\documentclass[10pt,a4paper]{article}
\usepackage[top=0.85in,left=1.6in,footskip=0.75in]{geometry}
\usepackage{amsmath,amssymb}
\usepackage{changepage}
\usepackage[shortlabels]{enumitem}

\usepackage[utf8x]{inputenc}

\usepackage{textcomp,marvosym}

\usepackage{cite}

\usepackage{nameref}
\usepackage{hyperref}

\usepackage[right]{lineno}

\usepackage{microtype}
\DisableLigatures[f]{encoding = *, family = * }

\usepackage[table]{xcolor}

\usepackage{array}

\usepackage{tabularx}
\usepackage{caption}
\usepackage{booktabs}

\usepackage{soul,color}   
\usepackage{float}            
\usepackage{cleveref}      
\usepackage{multirow}      
\usepackage {tabu}			   
\usepackage{longtable}    
\newcolumntype{C}[1]{>{\centering\let\newline\\\arraybackslash\hspace{0pt}}m{#1}}
\usepackage{graphicx}     
\usepackage{subcaption} 
\usepackage{wrapfig}       
\usepackage{soul}            
\usepackage{pdfpages}  
\usepackage{hyperref}  
\usepackage{xcolor}      
\usepackage{glossaries} 
\usepackage[strings]{underscore} 

\usepackage{ifthen}
\usepackage{helvet}
\makeatletter
\newcommand{\showfont}{encoding: \f@encoding{},
  family: \f@family{},
  series: \f@series{},
  shape: \f@shape{},
  size: \f@size{}
}

\newcolumntype{+}{!{\vrule width 2pt}}

\newlength\savedwidth

\usepackage[aboveskip=1pt,labelfont=bf,font=small,labelsep=period,singlelinecheck=off]{caption}

\makeatletter
\renewcommand{\@biblabel}[1]{\quad#1.}
\makeatother

\usepackage{lastpage,fancyhdr,graphicx}
\usepackage{epstopdf}
\newacronym{stcc}{$\tau$}{step-wise target control centrality}
\newacronym{lti}{LTI}{linear time invariant}
\newacronym{fmri}{fMRI}{functional magnetic resonance imaging}
\newacronym{pbh}{PBH}{Popov-Belevitch-Hautus}
\newacronym{mds}{MDS}{minimum driver node set}
\newacronym{ms}{MS}{multiple sclerosis}
\newacronym[shortplural={HC}]{hc}{HC}{healthy control}
\newacronym[longplural={regions of interest}]{roi}{ROI}{region of interest}
\newacronym{dti}{DTI}{diffusion tensor imaging}
\newacronym{lag}{LAG}{left angular gyrus}
\newacronym{dmn}{DMN}{default mode network}

\begin{document}
\vspace*{0.2in}

\begin{flushleft}
{\Large
\textbf\newline{\textbf{The impact of aging on human brain network target controllability}\\
}
}

\vspace{5mm}


Giulia Bassignana\textsuperscript{1,2},
Giordano Lacidogna\textsuperscript{3},
Paolo Bartolomeo\textsuperscript{1}
Olivier Colliot\textsuperscript{1,2},
Fabrizio De Vico Fallani\textsuperscript{1,2,*}
\\
\bigskip
\textbf{1} Sorbonne Universites, Paris Brain Institute (ICM), Inserm, CNRS, APHP, Hopital  Pitie Salpetriere, Paris, France \\
\textbf{2} Inria Paris, Aramis Project Team, Paris, France\\
\textbf{3} Hospital, Tor Vergata Unviersity, Rome, Italy
\\

\bigskip

%
%
%



* corresponding author: fabrizio.de-vico-fallani@inria.fr

\end{flushleft}
\bigskip


\subsection*{Abstract}
Understanding how few distributed areas can steer large-scale brain activity is a fundamental question that has practical implications, which range from inducing specific patterns of behavior to counteracting disease.

Recent endeavors based on network controllability provided fresh insights into the potential ability of single regions to influence whole brain dynamics through the underlying structural connectome.
However, controlling the entire brain activity is often unfeasible and might not always be necessary. The question whether single areas can control specific target subsystems remains crucial, albeit still poorly explored. \textcolor{black}{Furthermore, the structure of the brain network exhibits progressive changes across the lifespan, but little is known about the possible consequences in the controllability properties.}

To address these questions, we adopted a novel target controllability approach that quantifies the centrality of brain nodes in controlling specific target anatomo-functional systems. We then studied such target control centrality in human connectomes obtained from healthy individuals aged from $5$ to $85$.
Main results showed that the sensorimotor system has a high influencing capacity, but it is difficult for other areas to influence it. 
Furthermore, we reported that target control centrality varies with age and that temporal-parietal regions, whose cortical thinning is crucial in dementia-related diseases, exhibit lower values in older people.
By simulating targeted attacks, such as those occurring in focal stroke, we showed that the ipsilesional hemisphere is the most affected one regardless of the damaged area. Notably, such degradation in target control centrality was more evident in younger people, thus supporting early-vulnerability hypotheses after stroke.

\newpage
\section{Introduction}
 


Brain controllability refers to the possibility to induce specific functional states, or configurations, by means of internal or external control.
In neuroscience, this capability is associated with \textit{cognitive control} \cite{botvinick_motivation_2015}, which can be qualitatively assessed by measuring dynamic cooperation and competition between different neural systems during goal-directed tasks \cite{cocchi_dynamic_2013}.

Recently, the adoption of a control theoretic perspective has started to provide 
quantitative insights on how functional brain states can be predicted by the underlying brain network structure\cite{gu_controllability_2015}.
In particular, network controllability - i.e., the theoretical ability to guide a system's state by operating on few driver nodes - has received a growing interest in several biological applications \cite{yan_network_2017, ravindran_identification_2017, ravindran_network_2019, vinayagam_controllability_2016}.
Controllability of brain networks has been specifically explored to understand how the brain is able to endogenously modify its dynamics and if it is possible to steer it from exogenus stimuli \cite{tang_colloquium_2018, muldoon_stimulation-based_2016}.
Hence, a crucial step in network controllability is to determine the nodes that are the best-suited to drive the system's state.
To this end, centrality measures based on controllability were developed and applied to brain networks for identifying \textcolor{black}{potential driver nodes and measuring the size of the network that they can control  \cite{gu_controllability_2015, pasqualetti_controllability_2014}. }
These studies can have practical implications because they can inform possible intervention strategies to favor specific patterns of behavior or treat brain diseases, by means of brain stimulation technology \cite{bikson_safety_2016,wilson_clustered_2015,tang_colloquium_2018}.

While promising, controllability of the brain is still an underexplored 
research field due to the different ways in which controllability can be implemented \cite{jiang_irrelevance_2019}.
In practice, controlling the whole system from a single node could not {\color{black}{be feasible}} due to \textcolor{black}{algorithmic imprecision}  even in relatively small networks of hundreds of nodes. Furthermore, the \textcolor{black}{magnitude of the control signal might be too high to be generated physically or destructive for the network functioning itself \cite{suweis_brain_2019, tu_warnings_2018}. 
A possible solution would be to focus on specific parts of the network so as to reduce the overall computational complexity \cite{yan_network_2017,gao_target_2014, chen_optimizing_2020}.}  
Determining how single nodes can affect parts of the network still remains poorly understood in the human brain. Furthermore, current results have neglected the fact that brain controllability could vary with aging, which induces progressive changes in the underlying network structure or with focal brain damages \cite{zhao_age-related_2015, gong_age-_2009}.

\begin{figure}[ht!]
	\centering
	\includegraphics[width=\textwidth]{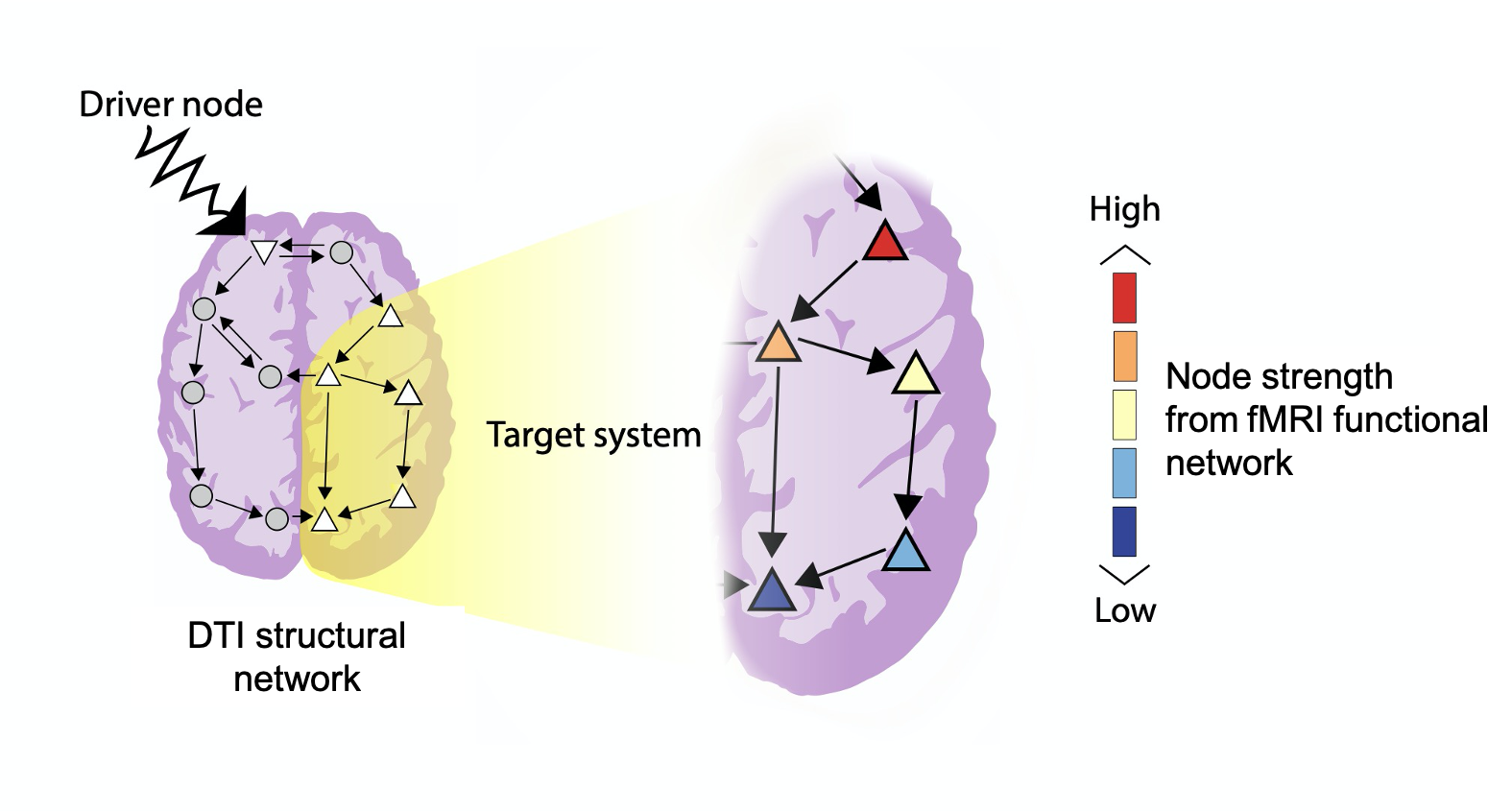}
	\caption{\small \textbf{Illustrating principle of the step-wise target control centrality.}
Left panel: the lower triangle indicates the selected driver node, upper triangles indicate the nodes in the chosen target system. The structural brain network derived from DTI data is the connection matrix of the LTI control system.
Right panel: the step-wise algorithm identifies the controllable driver-target configuration by visiting the target nodes \color{black}{according to their strength in the fMRI functional network}, i.e. from the highest to the lowest. 
}
\label{fig: stcc-scheme}
\end{figure}

To address these questions, we used cross-sectional anatomical $DTI$ and functional {\color{black}{\textit{MRI} (f\textit{MRI})}} data from {\color{black}{$171$}} healthy individuals aged between $5$ and $85$ \cite{brown_connected_2016}. 
Then, we evaluated the ability of single brain nodes to control specific target subsystems, i.e.,  frontal, limbic, temporal, sensorimotor, parietal, and occipital systems (\textbf{Fig. S1}).  To do so, we adopted an exact network controllability approach \cite{bassignana_step-wise_2020}, which gives for each tested driver the number of nodes in a predetermined target that it can control (\textbf{\autoref{fig: stcc-scheme}}).

We then studied how control centrality is distributed in the brain and which are the most controllable target systems. We assessed how control centrality is altered by age and whether there are specific regions that better explain the aging process. Finally, we tested how control centrality changes in presence of simulated targeted attacks and how the overall damage depends on the {\color{black}{participant}}'s age.

\section{Results}

\subsection{Sensorimotor areas have high control centrality}

\begin{figure}[ht!]
	\centering
	\includegraphics[width=13.5cm]{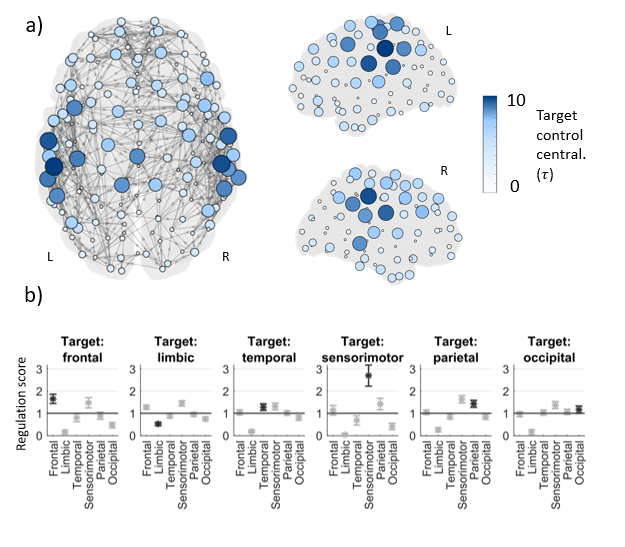}
	\caption{\small \textbf{Group-averaged target control centrality and \color{red}{regulation score}.}
	Panel a) cortical maps of $\tau$ values averaged across different target systems. Top view on the left and lateral views on the right. L=left hemisphere, R=right hemisphere.
	Panel b) {\color{black}{Regulation}} scores between different systems (\textbf{Methods}). Dark grey markers denote group-averaged {\color{black}{self-regulation scores}}, i.e. the ability of a system to control itself.  \textcolor{black}{Light grey markers denote group-averaged regulation scores between different systems and are reported for the sake of completeness.} 
	Error bars stand for standard deviation across {\color{black}{participants}}. 
}
\label{fig: br-centrality}
\end{figure}

We first studied how target control centrality is distributed across different functional systems of the human brain (\textbf{\autoref{fig: br-centrality}}, Methods).
Namely, we analyzed the ability of a single node in controlling different cortical systems by computing the target control centrality $\tau$, {\color{black}{the number of target nodes that this driver can control}} (\textbf{\nameref{sec: methods}}).
The group-averaged results showed that $\tau$ values significantly vary across regions (one-way ANOVA, $df = 5,\, F>969,\, p<10^{-6}$) as well as depending on the targeted system.
In particular, the sensorimotor regions had a high tendency to control the nodes in other systems, but they were also difficult to control from other areas (\textbf{Fig S2}). 
By pooling the results obtained with all different target systems, it became clearer that the sensorimotor nodes have on average the largest centrality values, while the limbic nodes have the lowest ones (\textbf{\autoref{fig: br-centrality}a}).

Next, we evaluated the tendency of single brain areas to control the system they belong to as compared to other systems.
To this end, we calculated the \textit{self-regulation} score of a node as the ratio of $\tau$ when targeting its own system and the average obtained when it targets all the other systems (\textbf{\nameref{sec: methods}}).
Results showed that on average brain regions tend to have a high self-regulation so that they control more nodes in their own system than those in other systems (one-way ANOVA, $df=4,\, F>1405, \,  p<10^{-6}$). 
That was particularly true for the sensorimotor system which consistently presented the highest values of the self-regulation score. On the contrary, the nodes in the limbic system exhibited a very low self-regulation (\textbf{\autoref{fig: br-centrality}b}).

\subsection{Control centrality decreases with age}

By exploiting the high variability in the age of the {\color{black}{participants}} (mean $35.8$ years, std $20.0$ years), we asked whether and how target control centrality varied with age.   
To this end, we computed the Spearman partial correlation between the $\tau$ values and age of the {\color{black}{participants}}, correcting for the outdegree of the nodes.
This correction was needed to exclude the existence of significant associations merely due to the presence of nodes with a  high number of outgoing links.
Results show a general tendency of the brain areas to be negatively correlated ($|R|> 0.15$,$p<0.05$, \textcolor{black}{FDR-corrected for multiple comparisons}) with age regardless of the targeted system (\textbf{Fig. S3}). 
\textcolor{black}{The presence of positive correlations is instead less consistent and weakly concentrated in the frontal and central areas of the brain}.

\begin{figure}[ht!]
	\centering
	\includegraphics[width=10cm]{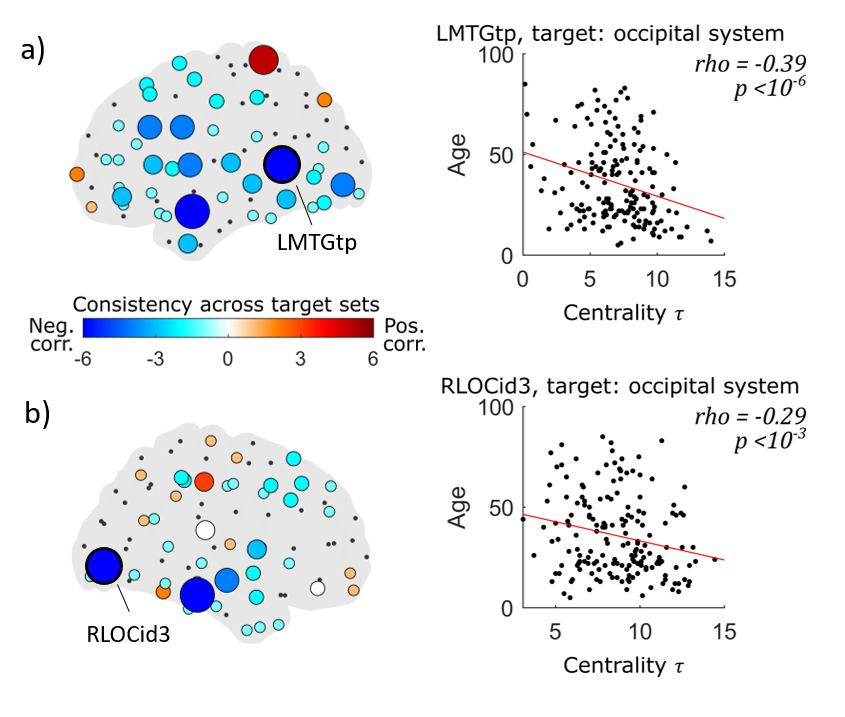}
	\caption{\small \textbf{Statistical correlation between target control centrality and age.} Panels a) and b) show respectively the correlation maps for the left and right hemisphere of the brain. The size and color of nodes code the number of times that a node exhibit a significant ($p<0.05$) correlation (negative-blue or positive-red) for different target systems.
Statistical correlation is computed by a partial Spearman correlation corrected for the out-degrees of the nodes. The insets on the right show the population scatter plots for the ROIs with highest correlation in each hemisphere. 
}
\label{fig: br-part-corr}
\end{figure}

By focusing on the ROIs that presented a significant association with the same sign across all targets, four regions emerged with a consistent negative correlation with age (\textbf{\autoref{fig: br-part-corr}}).
\textcolor{black}{Among those, the right lateral occipital inferior area (RLOCid3) in the occipital system is known to be associated with attentional processes and related to the dorsal attention network (\textbf{\autoref{fig: br-part-corr}b})\cite{zhang_distinct_2019}.}
The other regions were in the temporal system and three of them ({\color{black}{left middle temporal anterior gyrus (}}LMTGad{\color{black}{), left middle temporal temporo-occipital gyrus (}}LMTGtp{\color{black}{), right middle temporal posterior gyrus (}}RMTGpd{\color{black}{)}} belonged to the \gls{dmn} \cite{xu_activation_2016, davey_exploring_2016}.
The left middle temporal {\color{black}{temporo-occipital}} gyrus (LMTGtp) had the strongest negative correlation, possibly related to hippocampal degeneration with age \cite{scheltens_atrophy_1992}(\textbf{\autoref{fig: br-part-corr}a}). 
Taken together, these results suggested that most brain regions, and in particular those located in the temporal system, tend to decline with age.

\subsection{Targeted attacks lead to greater control centrality loss in younger brains}

Finally, we asked to what extent the target control centrality was impacted by attacks to specific brain systems, like those occuring after stroke, traumatic brain injury or tumor resection \cite{charras_functional_2015, salvalaggio_post-stroke_2020}.
To answer this question, we simulated lesions to different target systems by removing the nodes and the links from only one hemisphere. 
Then we evaluated the ability of all the other nodes to control the contralesional part of the target in the intact hemisphere, and we computed the difference with the original values ($\Delta_{attack}=\tau_{lesion}-\tau$) to quantify the impact of the damage.

\begin{figure}[ht!]
	\centering
	\includegraphics[width=\textwidth]{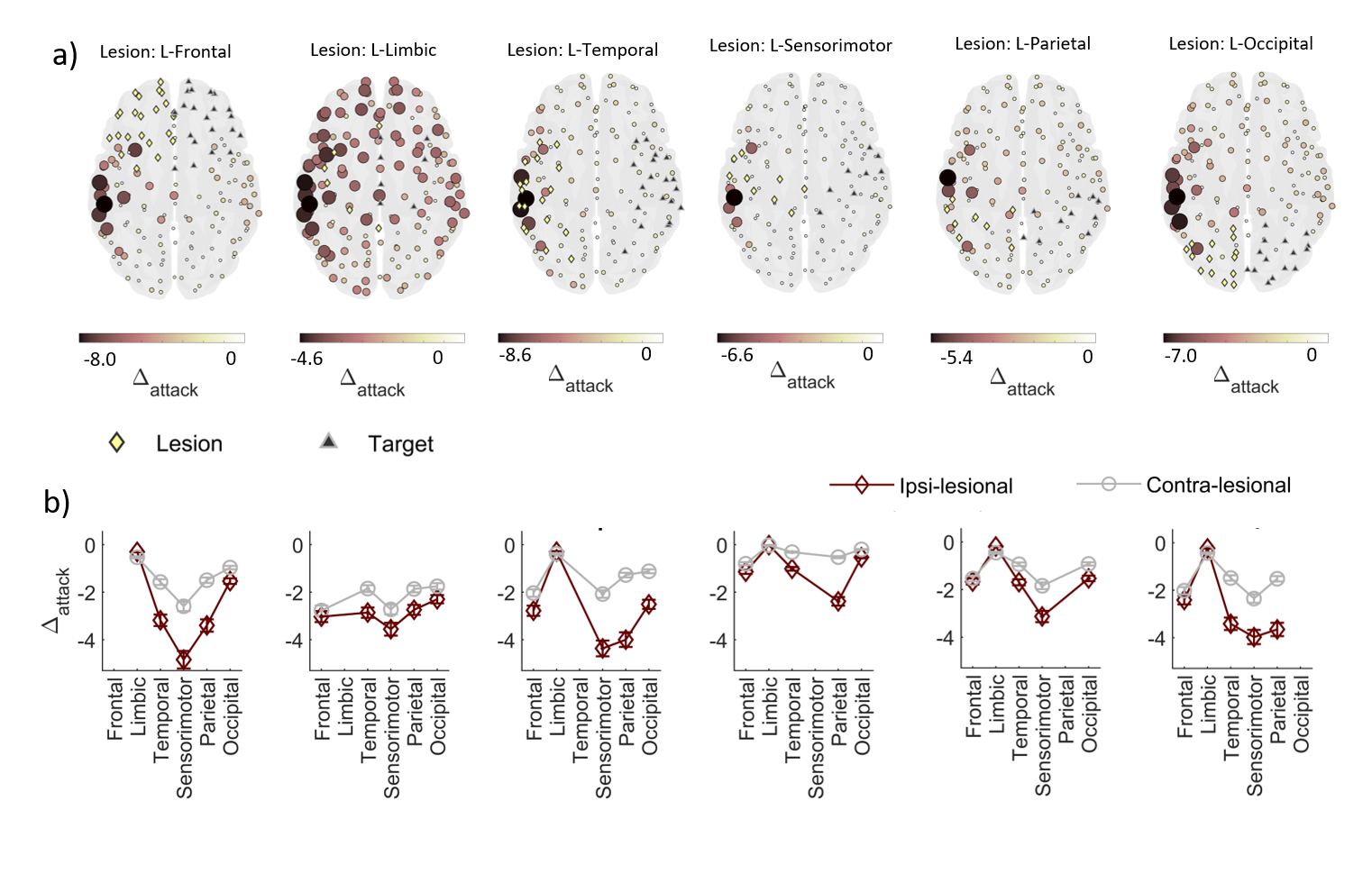}
	\caption{\small \textbf{Loss of target control centrality after simulated attacks.}
Panel a) shows the cortical maps of the group-averaged $\Delta_{attack}$ values when attacking different target systems on the left hemisphere (see \textbf{Fig. S4} for attacks to the right hemisphere). Size and color of the nodes are proportional to the decrease of target control centrality with respect to the \color{black}{healthy} situation.
Panel b) shows the values of the $\Delta_{attack}$ averaged across the nodes of a same system. Both values for the systems in the ipsilesional and contralesional hemisphere are illustrated.
Error bars stand for the standard error across \color{black}{participants}. 
}
\label{fig: br-lesion-L}
\end{figure}

\begin{figure}[ht!]
	\centering
	\includegraphics[width=10cm]{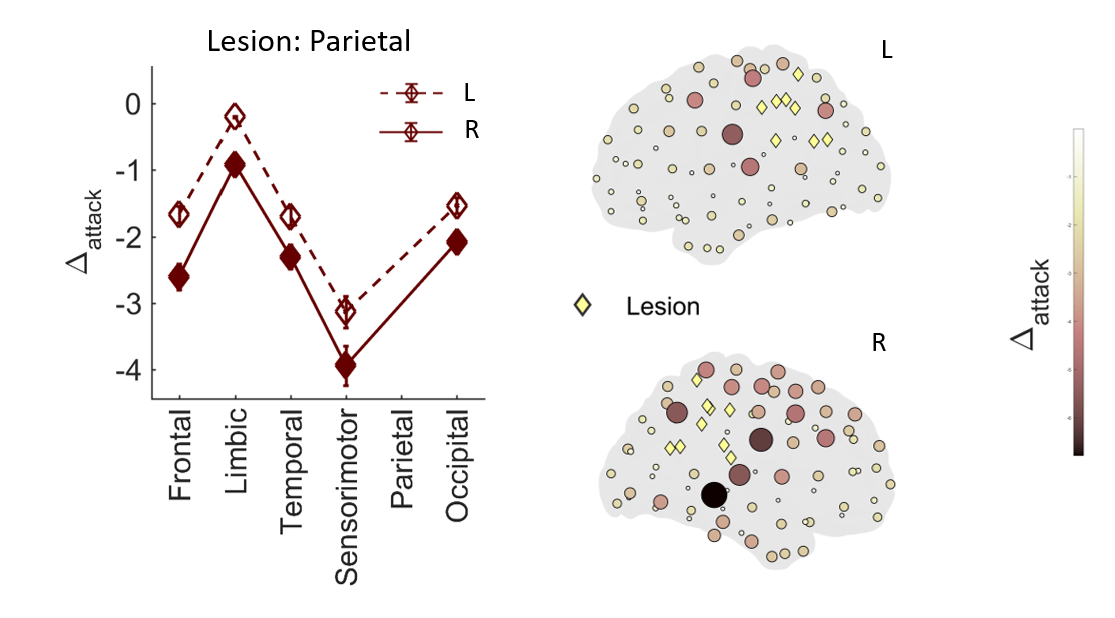}
	\caption{
	\small \textbf{Comparison of target control centrality loss between lesions to the right and left parietal system.}
	Left side panel illustrates the \color{black}{group-averaged} values for the $\Delta_{attack}$ of the intact systems when attacking the ipsilateral parietal system. Error bars stand for the standard error across \color{black}{participants}. Right panels show the cortical maps of the group-averaged $\Delta_{attack}$ values when attacking respectively the parietal system in the left (L) and right (R) hemisphere.
	}
\label{fig: br-lesion-parietal}
\end{figure}

As expected, network attacks led to decreases of control centrality, with greater losses in the ipsilesional hemisphere as compared to the unaffected one (two-ways ANOVA, $df=4, \, F>15, \, p<10^{-6}$, \textbf{\autoref{fig: br-lesion-L}} , \textbf{Fig. S4}).
{\color{black}{In particular}}, results showed that control centrality losses $\Delta_{attack}$ were globally small when the sensorimotor system was lesioned. Instead, when other systems were damaged, the sensorimotor system exhibited larger $\Delta_{attack}$ decrements. Conversely, the limbic system was relatively mildly impacted by any attack. 
In terms of difference between hemispheres we did not report a clear pattern across damaged systems. However, we consistently observed a greater ipsilesional centrality loss when the parietal system was attacked on the right hemisphere (two-ways ANOVA, $df=4, \, F>204, \, p<10^{-6}$, \textbf{\autoref{fig: br-lesion-parietal}}), which would reflect the the ability of the right parietal lobe to be involved in several high order cognitive processes \cite{corbetta_reorienting_2008, husain_space_2007, bartolomeo_hemispheric_2019} and interacting more with the other hemisphere as compared to the left one \cite{gotts_two_2013, koch_asymmetry_2011}.

\begin{figure}[ht!]
	\centering
	\includegraphics[width=10cm]{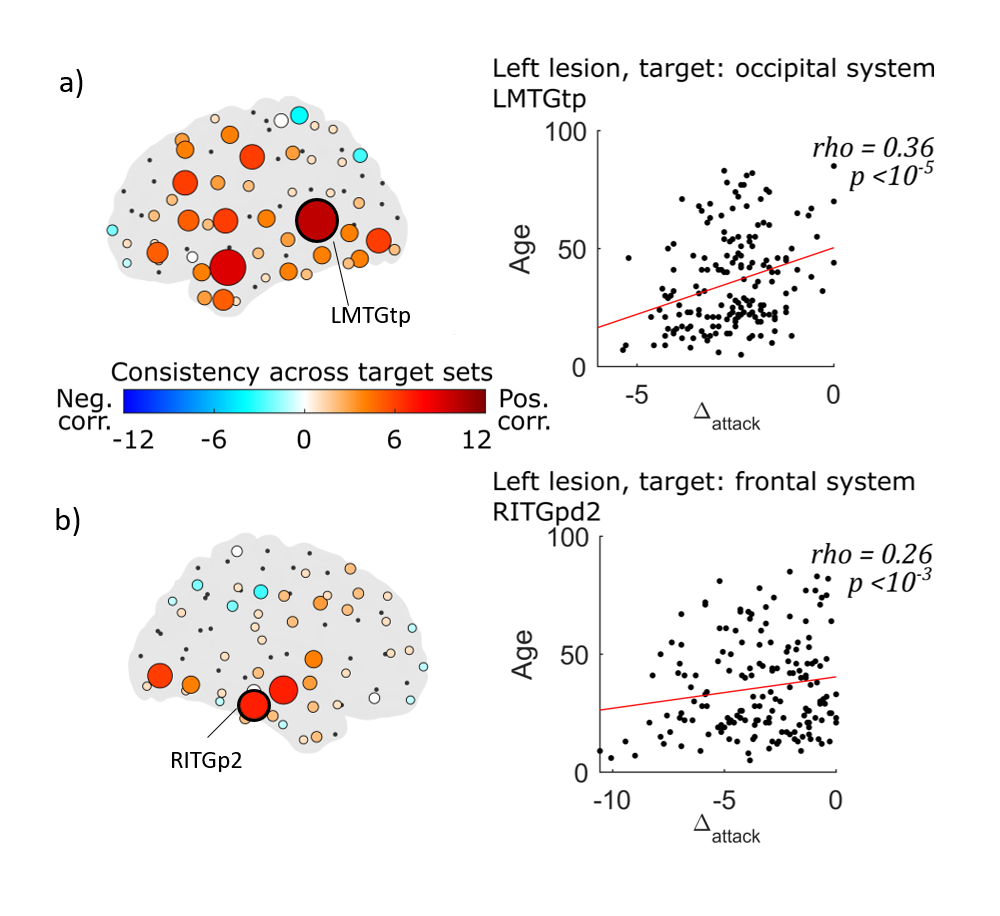}
	\caption{\small \textbf{Statistical correlation between target control centrality loss ($\Delta_{attack}$) and age.} Panels a) and b) show respectively the correlation maps for the left and right hemisphere of the brain. The size and color of nodes code the number of times that a node exhibit a significant ($p<0.05$) correlation (negative-blue or positive-red) for different target systems.
Statistical correlation is computed by a partial Spearman correlation corrected for the out-degrees of the nodes. The insets on the right show the population scatter plots for the ROIs with highest correlation in each hemisphere. 
}
\label{fig: corr stroke}
\end{figure}

Finally, we evaluated whether the age of the {\color{black}{participants}} had an impact on the observed node control centrality loss after the simulated network attacks. Results showed a global positive correlation between $\Delta_{attack}$ values and age (Spearman correlation, $|R|>0.15$, $p<0.05$, \textcolor{black}{FDR-corrected for multiple comparisons}), \textbf{\autoref{fig: corr stroke}}, \textbf{Fig. S5}), suggesting a more important effect in younger {\color{black}{participants}} compared to adults.
Notably, these  associations were consistently reported in the same temporal DMN areas, for which we also observed a significant positive correlation between control centrality values and age (\textbf{\autoref{fig: br-part-corr}}).


\section{Discussion}
\label{subsec: br-discussion}


\subsection{Driver nodes in the human brain}

Thanks to its ability to establish theoretical relationship between structure and dynamics, control network theory has been increasingly adopted to study human brain functioning. 
Control centrality, measuring how important a network node is in influencing other nodes' states, has been shown to provide useful markers of brain dynamics during resting states \cite{gu_controllability_2015} and executive functioning \cite{cui_optimization_2020}, as well as to predict responses to brain stimulation in practice \cite{medaglia_network_2018} and in theory \cite{muldoon_stimulation-based_2016}.
These findings establish an intuitive link between the brain regions identified as potential control drivers and the current knowledge in cognitive neuroscience. 
Notably, large-scale brain activity has been found deeply dependent on nodal controllability within specific systems, such as default mode, fronto-parietal, cingulo-opercular, and attention \cite{honey_can_2010}. 

While these results provide mechanistic insights on how single areas can steer whole-brain activity, they do not inform on how different specialized brain systems can influence each other. Is there a dominant subnetwork prone to control all the other ones? Is there a subsystem that is instead more controllable than others? 
These fundamental questions become particularly pressing considering that one-to-whole control might be difficult to achieve both from a theoretical and practical angle \cite{suweis_brain_2019, tu_warnings_2018}.
To answer those questions, we used the step-wise target control centrality, i.e. an optimized Kalman-based heuristics that measures the ability of nodes to control predetermined target parts of the network \cite{bassignana_step-wise_2020}.

Our results showed that in general node control centrality varies depending on the targeted system. Despite such variability, the sensorimotor areas always exhibited the highest influencing capacity while being very hard to control. Conversely, the limbic regions always had a scarce driving power and they were on average easily controllable by all other nodes (\textbf{\autoref{fig: br-centrality}}).
This tendency was also confirmed when looking at the ability of those systems to control, or self-regulate, themselves.

The sensorimotor system is known to be more densely connected, with many anatomical long-distance interhemispheric connections, as compared to other secondary motor-related systems \cite{narayanan_redundancy_2005}. Such hyperconnectivity, would constitute therefore a structural prerequisite to reach and orchestrate different areas during cognitive and motor functions \cite{paquola_microstructural_2019, shafiei_topographic_2020, wang_macroscopic_2020, vazquez-rodriguez_gradients_2019, demirtas_hierarchical_2019}. 
From a functional perspective the high control centrality of the sensorimotor system has been previously associated with its ability to process information not only for motor control but also for a broad range of recognition processes \cite{sohn_network_2021, bassett_learning-induced_2015, adolphs_role_2000}.
This is in line with the existence of gradients of cortical organization through which the sensorimotor system could boost neural activity of other association areas required for higher order functions such as cognitive control, guided attention and motivation \cite{kong_sensory-motor_2021}. 
\textcolor{black}{In line with such a cortical organization gradient, we could have expected that the visual system -another highly connected primary system- would have exhibited high control centrality, too. However the control centrality of the nodes in the occipital lobe, where main hubs of the visual system are localized, were rather low. 
Future works will be crucial to assess whether the high control centrality of the sensorimotor system is mainly due to its high connection density and/or specific local synaptic properties and gradients of gene expression in the excitation–inhibition balance of related interneurons \cite{kong_sensory-motor_2021}.}

While limbic areas constitute an important structural bridge for the information transfer between cortical and subcortical regions \cite{rolls_limbic_2015, catani_revised_2013}, their control centrality was remarkably low regardless of the targeted cortical system. 
This could be in part explained by the heterogenous nature the lymbic system, which includes different anatomical and cytoarchitectonic components \cite{catani_revised_2013, kaas_evolution_1995, rasia-filho_subcortical-allocortical-_2021}.
However, it is also true that many important subcortical areas, such as amygdala, caudate nucleus, and hypothalamus \cite{rolls_limbic_2015}, were not available in the dataset we used (see \textbf{\nameref{subsec: methodological-considerations}} for more details). 
Future research will be crucial to elucidate the role of lymbic areas in terms of target control centrality using more comprehensive and accurate subcortical-cortical systems.

\subsection{Aging and control centrality}

Brain aging is a highly heterogeneous and dynamic process that involves structural and functional changes both at individual and group level. Among structural changes, cortical thinning and regional atrophy \cite{bakkour_effects_2013}, white matter loss of integrity \cite{damoiseaux_white_2009}, neuronal loss and degeneration \cite{allen_normal_2005} and neurotransmitters depletion have been detected at varying degrees among older subjects \cite{allen_normal_2005}.
Functional alterations refers to maladaptive, age-related brain activity, detected in neuroimaging studies, including decreased specificity of  ventral-visual and motor areas \cite{bernard_evidence_2012, voss_dedifferentiation_2008}, decreased memory-related recruitment of medial temporal lobe regions \cite{cabeza_task-independent_2004} and dysregulation of the default mode network \cite{park_adaptive_2009}.

Network approaches investigating brain reorganization across the lifespan have constantly reported whole-brain intrinsic connectivity changes by using standard centrality metrics such as node degree \cite{hampson_intrinsic_2012}, strength \cite{bagarinao_aging_2020} and betwenness \cite{bagarinao_reorganization_2019}. 
Compared to young adults, older ones exhibit reduced local-efficiency and modularity as well as connectivity changes within and between specific subnetworks \cite{geerligs_brain-wide_2015, betzel_changes_2014, song_age-related_2014, spreng_attenuated_2016}.
Only few studies have attempted to study how network controllability evolves with age. In a recent study, authors showed that regional controllability increases with development \cite{tang_developmental_2017}, but no information is available on how it evolves across the lifespan.
To address this question, we combined information from both structural (\acrshort{dti}) and functional (\acrshort{fmri}) neuroimaging data, and we analyzed the effect of aging on a new measure of centrality based on network controllability. 

Main results showed that control centrality was negatively correlated with age indicating a global trend of node controllability reduction in older brains. 
\textcolor{black}{The presence of sporadic positive correlations in the frontal and central areas should be further investigated for possible compensatory mechanisms occuring in later age as well as in mild cognitive impairment and Alzheimer's disease \cite{kubicki_early_2016, behfar_graph_2020, guillon_disrupted_2019}.}
Instead, brain regions in the middle temporal gyrus (MTG) were significantly impacted by aging, in terms of relative loss of control centrality (\textbf{\autoref{fig: br-part-corr}}). 
Those areas were previously found to be less activated in elderly people, reflecting a lower semantic retrieval control process \cite{davey_exploring_2016, hoffman_age-related_2018}.
Such a decrement could be associated with the the more general age-dependent DMN functional rearrangement \cite{grady_age_2016, li_neuromodulation_2014}. 

According to the ``network dedifferentiation'' hypothesis \cite{chan_decreased_2014}, DMN regions in older adults progressively present a reduction in the communication with other systems, such as the dorsal attention network (DAN) or the frontoparietal network\cite{spreng_attenuated_2016, de_schotten_direct_2005}.
Thus, age-related reduced control centrality could be a possible reorganizational mechanism subserving the failure to deactivate neural systems that are unrelated to the task \cite{grady_cognitive_2012} and leading to abnormal increased brain activation \cite{duda_neurocompensatory_2019, morcom_neural_2015} and negative correlation with task performance \cite{buckner_brains_2008, logan_under-recruitment_2002}.

\textcolor{black}{Our results showed that while there is a common distribution between the age-centrality correlations in the left and right hemisphere, there are also few notable differences.}
The specific asymmetric involvement of the right lateral occipital inferior cortex (RLOCid3) could be intriguingly related to a progressive inability to mediate the interaction between the right ventral attention system, involved in saliency analysis, and the bilateral dorsal attention system, mostly involved in attention shifting \cite{corbetta_control_2002, shulman_right_2010, bartolomeo_hemispheric_2019}. 
The interaction between the two attentional systems finally influences the processing of attended visual stimuli in the primary {\color{black}{visual}} cortex \cite{murray_attention_2004, de_schotten_direct_2005}.
While asymmetry patterns could be interpreted in the light of age dependent loss of lateralization in specific cognitive processes, such as semantic control \cite{grady_cognitive_2012, cabeza_hemispheric_2002}, more investigation is needed to better disentangle this aspect.

Finally, recent studies suggest that long-range connections may be more vulnerable to aging effects than short-range connections in both DMN and DAN \cite{tomasi_aging_2012}. A future perspective would be to check to what extent the observed controllability changes depend on the relative spatial distance between the driver and target nodes.

Overall, our results indicated that age related reduction in controllability in specific nodes at the crossroad of different brain systems relevant for cognitive control efficiency, could be viewed as a yet another evidence of a functional ``reshaping'' of brain networks along the lifespan \cite{spreng_shifting_2019, grady_cognitive_2012}. 


\subsection{Control centrality markers of brain lesions}

Network control theory not only informs on the basic brain functioning but it can also offer new analytical ways for quantifying dysfunctions in neurological diseases \cite{medaglia_brain_2017}. 
After a focal damage, such as in stroke, traumatic injuries and tumors, the brain typically loses the functions associated with the lesioned area and with those connected to it. 
Quantifying the effects of such local destruction on the rest of the network is therefore crucial to predict the extent to which the brain will recover its functions through a reorganizational process, i.e. \textit{plasticity} \cite{cheng_reorganization_2012, zhu_disrupted_2017}.

Here, we studied how control centrality changed when simulating unilateral attacks to the target systems.
Results showed that the nodes in the ipsilesional hemisphere globally underwent larger losses in control centrality. That was particularly true for the sensorimotor regions regardless of the system attacked.
From a mechanistic perspective, the ability of the nodes in the damaged hemisphere to control the unaffected side of the target, would be compromised by the removal of the interhemispheric homotopic connections, i.e. the links bridging the homologous regions in the two hemispheres \cite{mancuso_homotopic_2019, tang_decreased_2016}.
The interruption of such homotopic bridge, which is typically stronger compared to heterotopic connectivity, eventually made the nodes in the intact target less reachable from the nodes in the lesioned hemisphere. 
\textcolor{black}{The extent to which this also affects possible compensatory actions from the contralesional ``intact'' hemisphere, remains a question to be elucidated with more clinical and longitudinal data \cite{grefkes_connectivity-based_2014, buetefisch_role_2015, bartolomeo_let_2016, bartolomeo_competition_2021, kullmann_editorial_2019}.}


While the controllability power of ipsilesional sensorimotor regions was particularly impacted when damaging the targets, the other regions appeared to be less affected  when the sensorimotor target was attacked. 
This weak effect was in part due to the relatively low values of control centrality of those regions, which limited the possible range of change after the damage. {\color{black}{More generally}}, this might be related to a putative high reachability of the sensorimotor system given its high connection density \cite{reich_independent_2001, narayanan_redundancy_2005, so_redundant_2012} and the presence of alternative pathways between frontal, parietal and limbic areas \cite{griffis_structural_2019, betzel_optimally_2016}.

We observed a progressive decrease of control centrality with age, mostly localized in the temporal lobes (\textbf{\autoref{fig: corr stroke}}). 
We then asked whether the control centrality losses induced by the lesions, regardless of the targeted system, could also depend on the age of individuals. 
Results confirmed that the damage-related control centrality reductions in the temporal lobes were larger in younger brains, which have in general higher baseline values of controllability. 

From a biological perspective, it’s well established that changes in neuroplasticity occur lifelong through many age-specific processes \cite{dosenbach_prediction_2010}, ultimately affecting cerebral network maturation \cite{fair_functional_2009, supekar_developmental_2012} and controllability \cite{tang_developmental_2017}. 
Likewise, brain lesions could impact brain organizational properties in an age-dependent manner. 
In this direction, two seemingly contradictory explanations have emerged: first, ``early plasticity'', arguing for the greater flexibility of the immature brain, and associated good recovery and outcome \cite{giza_is_2006, kornfeld_cortical_2015}; and secondly, ``early vulnerability'' referring to the young brain's unique susceptibility and subsequent poor outcome \cite{anderson_children_2011, max_pediatric_2010, dennis_age_2013}. 

Our results indicated that target control centrality follows an age-dependent trend that mirrors an early vulnerability condition, so that networks of younger {\color{black}{participants}} are more impacted by focal lesions, possibly reflecting a reduced resiliency to damages in the early stages. 
\textcolor{black}{Shedding a light on functional reorganization and recovery after brain injuries or stroke might enable better prediction and prevention of clinical outcome. However, the lack of clinical data for our population allows us neither to validate this result from a clinical perspective nor to assess the early-plasticity hypothesis which would require longitudinal clinical data associated with the recovery of patients.}

Future research is needed to better identify possibly longitudinal changes in network controllability related to specific clinical outcome in both earlier and later stages of brain reorganizaiton after brain damages \cite{anderson_children_2011}.

\subsection{Methodological considerations}
\label{subsec: methodological-considerations}
In this work, we exploited the step-wise target controllability framework \cite{bassignana_step-wise_2020}, which relies on a \acrfull{lti} dynamics and is based on the Kalman rank condition. 
While it is known that the brain presents a nonlinear dynamics, the study of linear models has proved to be beneficial in improving our understanding \cite{gu_controllability_2015, liu_control_2016, tang_colloquium_2018}.
Specifically, the controllability of a linearized model can inform on the controllability of the nonlinear model \cite{slotine_applied_1991}.

In order to study the controllability of brain networks efficiently, we gave a directionality to the connectomes.
Despite the fact that it is not currently possible for neuroimaging techniques to discern the directionality of bundles of axons, it is known that each neuron propagates signals through a well-defined direction from the soma to the axon terminal.
Previous efforts to direct a connectome relied on the hypothesis that, given a set of brain networks, edges present for all {\color{black}{participants}} are the oldest, and any new edge would be directed from the new node to the existing cluster \cite{kerepesi_how_2016}. 
However, a limitation  of this method is that it is strictly dependent on the initial set of networks, and the procedure would not be easily scalable.
Other methods based on local navigation and communicability were devised to infer the directionality of neural signaling, but since they operate best for nodes connected by longer paths, they are not well-suited to perform inference for structurally connected nodes \cite{seguin_inferring_2019}. 
Our approach was instead inspired by the diffusion processes taking place in the connectome \cite{worrell_optimized_2017, goni_exploring_2013, avena-koenigsberger_communication_2018, raj_network_2012, abdelnour_functional_2018} and preserves the hierarchical and modular properties of the anatomical pathways.

Finally, it is important to mention that the obtained results refers to the specific way we have selected the target systems and \textcolor{black}{and ranked the nodes therein contained}, and should not be generalized to other possible choices.
By adopting an ``anatomical proximity'' criterion, some regions that could be functionally associated to the frontal lobe, e.g. the anterior cingulate cortex ACC, have been instead assigned to the lymbic system.
The original dataset did not contain subcortical areas, such such as amygdala, caudate nucleus, and hypothalamus \cite{catani_revised_2013}, which are known to be tightly related to the lymbic system.
\textcolor{black}{Future research will therefore be important to validate the obtained results on different brain atlases and target system selection. Furthermore, more rigorous anatomical and circuit-level information from histological and causal experiments can better elucidate the robustness of our results obtained from noninvasive neuroimaging \cite{dubois_single-unit_2015, schilling_histological_2018}.}

\section{Conclusions}
In this study, we presented a method to quantify the ability of candidate driver nodes to drive the state of a target set in directed brain networks.
The obtained results revealed that sensorimotor areas are theoretically inclined to control different target systems, while regions in the temporal lobe were negatively impacted by age and by simulated damages to the network.
These results are in line with the general claim of a dominant gradient of cortical organization with sensory-motor and association regions at opposing ends \cite{huntenburg_large-scale_2018, malkinson_perception_2021}.
We hope that further developments of network controllability measures will contribute to the identification of the key nodes in biological networks to better identify targets of brain stimulation to counteract human diseases.

\section{Methods}
\label{sec: methods}
\paragraph{Structural and functional brain networks}
We used already processed brain network data from the NKI-Rockland database \cite{nooner_nki-rockland_2012}, and selected $171$ {\color{black}{participants}}, aged from $5$ to $85$ \cite{brown_ucla_2012}.  For each {\color{black}{participant}} we had access to both structural (\acrshort{dti}) and functional (\acrshort{fmri}) data.
\textcolor{black}{Both structural and functional networks consisted of $188$ nodes corresponding to functional regions of interest (ROIs) established with a spatially constrained spectral clustering method \cite{craddock_whole_2012}. All related information including node labels and spatial position can be accessed via the USC Multimodal Connectivity Database \url{http://umcd.humanconnectomeproject.org/umcd/}.}
\textcolor{black}{We also provided a compacted version of this information in a dedicated supplementary file (\textbf{File S1}). Note that some ROIs had been split in different parts. In this case, we added a suffix number at the end of the label to specify its relative position along the longitudinal axis of the brain (i.e. higher numbers, more caudal positions). }

\paragraph{Directing structural brain networks}
We directed the structural connectomes in order to study their controllability properties in a more efficient way, and we used the information from fMRI brain networks to establish a hierarchy among the nodes in the target systems.
Starting from the \acrshort{dti} network, first we applied a logarithmic transformation ($\log{(w+1)}$) to make the weights more homogeneous.
Then, we performed a biased random walk \cite{gomez-gardenes_entropy_2008} to direct each edge from the node with lower strength to the node with higher strength. The probability $P_{ij}$ to go from node $i$ to node $j$, can be computed as
\begin{equation}
	P_{ij} = \frac{w_{ij}}{ \sum_h {w_{ih}} }
\end{equation}
where $[w]_{ij}$ is the symmetric, weighted adjacency matrix of the structural network.
For each pair of nodes $i$ and $j$, we had two directed edges with weights $P_{ij}$ and $P_{ji}$.
We chose one direction by keeping only the highest probability (we kept both if they were equal), thus drastically reducing the presence of loops in the network.

\textcolor{black}{The choice of removing some links was inspired by the works on the minimum spanning tree (MST) procedure applied to brain networks \cite{tewarie_minimum_2015}. Similarly to our strategy, the MST filtering procedure eliminates any loop in the network, yet important information can be obtained on the brain network structure and function with implication in aging \cite{van_dellen_minimum_2018}.
Despite such background, our choice still remains a modeling strategy that favors information propagation towards higher degree nodes. Note that the resulting effect is in line with the evidence that hubs in real networks tend to have less control power (mechanistically exerted by outgoing links), as they instead need to be efficiently accessed to control the rest of the network \cite{liu_controllability_2011}.
As for the reduction of loops, it is important to stress that this might lead to underestimate the number of controllable nodes, but not to introduce false positive results.
This is because of the underlying k-walk theory by which a driver-target configuration is controllable if the length of the path from the driver to each target is unique \cite{gao_target_2014}. }

\paragraph{Step-wise target controllability}
We implemented the step-wise target controllability framework \cite{bassignana_step-wise_2020}, that analyses the single-input target controllability problem, in which the interest is to study the role of a single driver node in controlling a target set of the system.
We assumed the \acrfull{lti} dynamics
\begin{equation}
	\dot{\textbf{x}}(n) = A \textbf{x}(n) + B \textbf{u}(n), \qquad \textbf{y}(n) = C\textbf{x}(n)
	\label{eq: LTI}
\end{equation}

where $\textbf{x}(n) \in \mathbb{R}^N$ describes the state of each node at time $n$, $A$ is the adjacency matrix of the network, $B \in \mathbb{R}^N$ specifies the driver node that will receive an external input, $\textbf{u}(n) \in \mathbb{R}^N$ is its external input (or control) signal, $\textbf{y}(n) \in \mathbb{R}^S$ is the output vector, and $C \in \mathbb{R}^{S \times N}$ is the output matrix identifying the target nodes. 

Such a system is controllable if it can be guided from any initial state to any desired final state in finite time, with a suitable choice of input. 

{\color{black}{A necessary and sufficient condition to assess the controllability of \textbf{Eq. \ref{eq: LTI}}, is that the controllability matrix $Q$

\begin{equation}
	Q = \begin{bmatrix} B&AB&A^{2}B&\cdots &A^{N-1}B \end{bmatrix} 
	\label{eq: kalman matrix}
\end{equation}

has full row rank, i.e. rank$(Q)=N$. That is the Kalman rank condition, which basically verifies the existence of linearly independent rows in $Q$ \cite{kalman_mathematical_1963, rugh_linear_1995}. If so, the driver node can reach and control the dynamics of all the other nodes through independent walks of length $N-1$ at maximum.

If it is of interest to control only a target set $\mathcal{T}$ of the network, specified in $C$ and consisting of $S \le N$ nodes, then \textbf{Eq. \ref{eq: LTI}} can be reduced into a target controllability matrix $Q_\mathcal{T} = C Q$, where $C$ filters the rows of interest corresponding to the targets. 
Now, the rank of $Q_\mathcal{T}$ gives the number $\tau \leq S$ of nodes in the target set that can be controlled by the driver.

To identify a driver-target configuration, we further introduce a hierarchy among the target nodes, so that we can order and relabel them from the most important one to the least,  i.e. $t_1 \succ t_2 \succ ... \succ t_S$.
Then we perform the following step-wise procedure for each candidate driver node.

\begin{itemize}

	\item Step 1. \textit{Initialization}
	
	\begin{itemize}
		\item Create a temporary empty target set $\mathcal{T'} \leftarrow \lbrace \rbrace$ 
		\item Set the number of controllable targets $\tau \leftarrow 0$
	\end{itemize}
	
	\item Step 2. \textit{Repeat until termination criteria are met.} For $j \leftarrow 1,...,S$ do
	
	\begin{itemize}
		\item Add the $j$-th target node to the target set $\mathcal{T'} \leftarrow \mathcal{T'} \cup \{ t_j \}$ 
		\item Build the subgraph containing the nodes on walks from the driver to the targets in $\mathcal{T'}$
		\item Compute the rank of the target controllability matrix $Q_\mathcal{T'}$
		\item \textit{If} rank($Q_\mathcal{T'}$) is full then $\tau \leftarrow \tau + 1$ else $\mathcal{T'} \leftarrow \mathcal{T'} \setminus \{t_j \}$
		\item $j\leftarrow j+1$
	\end{itemize}
	\item Step 3. \textit{Output $\tau$ and $\mathcal{T'}$} 
\end{itemize}

Eventually, the \textit{target control centrality} $\tau$ is the number of controllable targets in $\mathcal{T}$, and the set $\mathcal{T}'$ contains the $\tau$ controllable targets with highest ranking.

}}

\paragraph{\color{black}{Application to brain networks}}
In this specific application, 
\textcolor{black}{we assumed that the states of the nodes/ROIs were influenced by the adjacency matrix corresponding to the directed structural connectome. The target sets were the structural systems (frontal, limbic, temporal, sensorimotor, parietal, occipital, \textbf{Fig. S1}). 
\acrshortpl{roi} in the target system were ranked according to the group-averaged node strength obtained from the fMRI functional brain network (\textbf{\autoref{fig: stcc-scheme}}, \textbf{File S1})}.

\textcolor{black}{More precisely, we ranked the nodes in the target set $\mathcal{T}$ in a descending order from the quantity
$\frac{1}{M}\sum_m s_{i}^{(m)}$, where $s_{i}^{(m)}$ is the node strength (i.e. the weighted sum of all the connections) of the node $i$ obtained from the fMRI functional network of the participant $m$, and $M$ is the number of participants in the study. Node strengths quantify the tendency of brain areas to act as hubs which are crucial constituents of the overall information integration \cite{van_den_heuvel_network_2013}}.

The target control centrality \textcolor{black}{\textbf{$\tau$}} of each node is obtained via the stepwise procedure described above and gives the corresponding highest-ranked controllable configuration of target nodes.
To {\textcolor{black}{obtain}} a more robust estimation, we derived an intergated measure of target control centrality by averaging the results from the connectomes thresholded with different connection densities \cite{de_vico_fallani_graph_2014}. 
Specifically, for each {\color{black}{participant}}, we retained the strongest links so to have networks with mean node degree $k$ ranging from $1$ to $14$. This upper limit corresponded to the lowest number of links found after directing the connectomes.


\paragraph{{\color{black}{System regulation}} score}

\textcolor{black}{To quantify how the target control centrality of a system is globally distributed across the other target systems we introduced the so-called system regulation score $\mathcal{R}$.
Given two brain systems $i$ and $j$ the regulation score reads as}

\begin{equation}
	\mathcal{R}_{ij}=\frac{\frac{1}{|\mathcal{S}^{(i)}|}\sum_{k\in\mathcal{S}^{(i)}}\tau^{(j)}_k}{\frac{1}{N}\sum_{k \in \mathcal{V}}\tau^{(j)}_k} 
\end{equation} 

where $\mathcal{S}^{(i)}$ is the reference system $i$, $\mathcal{V}$ is the set of all $N$ nodes in the brain network, and $\tau_k^{(j)}$ is the target control centrality of each node in the system $i$ over the target system $j$.
When $i=j$, we specifically talk about self-regulation.

\textcolor{black}{\paragraph{Statistical analysis} 
All statistical analysis were performed with a statistical threshold of $p=0.05$, adjusted via a false-discovery rate (FDR) procedure in the case of multiple tests \cite{benjamini_controlling_1995}.}

\section*{Funding and/or Conflicts of interests/Competing interests}
Authors would like to acknowledge Thibault Rolland (fr.linkedin.com/in/thibault-rolland-40b57419a) for the realization of Picture 1. The research leading to these results has received funding from the French government under management of Agence Nationale de la Recherche as part of the "Investissements d'avenir" program, reference ANR-19-P3IA-0001 (PRAIRIE 3IA Institute) and reference ANR-10-IAIHU-06 (Agence Nationale de la Recherche-10-IA Institut Hospitalo-Universitaire-6), and from the Inria Project Lab Program (project Neuromarkers), the European Research Council (ERC) under the European Unions Horizon 2020 research and innovation programme (Grant Agreement No. 864729).
The content is solely the responsibility of the authors and does not necessarily represent the official views of any of the funding agencies.
Authors declare no conflict of interest.

\section*{Data availability statement}
All the experimental data used in this work are fully accessible from the NKI-Rockland database \cite{nooner_nki-rockland_2012}.

\nolinenumbers

\clearpage
\bibliographystyle{plos2015.bst}
\nolinenumbers

\begin{small}
\bibliography{Exported-Items.bib}

\begin{thebibliography}{100}

\bibitem{botvinick_motivation_2015}
Botvinick M, Braver T.
\newblock Motivation and {{Cognitive Control}}: {{From Behavior}} to {{Neural
  Mechanism}}.
\newblock Annual Review of Psychology. 2015;66(1):83--113.
\newblock doi:{10.1146/annurev-psych-010814-015044}.

\bibitem{cocchi_dynamic_2013}
Cocchi L, Zalesky A, Fornito A, Mattingley JB.
\newblock Dynamic Cooperation and Competition between Brain Systems during
  Cognitive Control.
\newblock Trends in Cognitive Sciences. 2013;17(10):493--501.
\newblock doi:{10.1016/j.tics.2013.08.006}.

\bibitem{gu_controllability_2015}
Gu S, Pasqualetti F, Cieslak M, Telesford QK, Yu AB, Kahn AE, et~al.
\newblock Controllability of Structural Brain Networks.
\newblock Nature Communications. 2015;6:8414.
\newblock doi:{10.1038/ncomms9414}.

\bibitem{yan_network_2017}
Yan G, V{\'e}rtes PE, Towlson EK, Chew YL, Walker DS, Schafer WR, et~al.
\newblock Network Control Principles Predict Neuron Function in the
  {{Caenorhabditis}} Elegans Connectome.
\newblock Nature. 2017;advance online publication.
\newblock doi:{10.1038/nature24056}.

\bibitem{ravindran_identification_2017}
Ravindran V, V S, Bagler G.
\newblock Identification of Critical Regulatory Genes in Cancer Signaling
  Network Using Controllability Analysis.
\newblock Physica A: Statistical Mechanics and its Applications.
  2017;474:134--143.
\newblock doi:{10.1016/j.physa.2017.01.059}.

\bibitem{ravindran_network_2019}
Ravindran V, Nacher JC, Akutsu T, Ishitsuka M, Osadcenco A, Sunitha V, et~al.
\newblock Network Controllability Analysis of Intracellular Signalling Reveals
  Viruses Are Actively Controlling Molecular Systems.
\newblock Scientific Reports. 2019;9(1):2066.
\newblock doi:{10.1038/s41598-018-38224-9}.

\bibitem{vinayagam_controllability_2016}
Vinayagam A, Gibson TE, Lee HJ, Yilmazel B, Roesel C, Hu Y, et~al.
\newblock Controllability Analysis of the Directed Human Protein Interaction
  Network Identifies Disease Genes and Drug Targets.
\newblock Proceedings of the National Academy of Sciences. 2016; p. 201603992.
\newblock doi:{10.1073/pnas.1603992113}.

\bibitem{tang_colloquium_2018}
Tang E, Bassett DS.
\newblock Colloquium: {{Control}} of Dynamics in Brain Networks.
\newblock Reviews of Modern Physics. 2018;90(3):031003.
\newblock doi:{10.1103/RevModPhys.90.031003}.

\bibitem{muldoon_stimulation-based_2016}
Muldoon SF, Pasqualetti F, Gu S, Cieslak M, Grafton ST, Vettel JM, et~al.
\newblock Stimulation-{{Based Control}} of {{Dynamic Brain Networks}}.
\newblock PLOS Computational Biology. 2016;12(9):e1005076.
\newblock doi:{10.1371/journal.pcbi.1005076}.

\bibitem{pasqualetti_controllability_2014}
Pasqualetti F, Zampieri S, Bullo F.
\newblock Controllability {{Metrics}}, {{Limitations}} and {{Algorithms}} for
  {{Complex Networks}}.
\newblock IEEE Transactions on Control of Network Systems. 2014;1(1):40--52.
\newblock doi:{10.1109/TCNS.2014.2310254}.

\bibitem{bikson_safety_2016}
Bikson M, Grossman P, Thomas C, Zannou AL, Jiang J, Adnan T, et~al.
\newblock Safety of {{Transcranial Direct Current Stimulation}}: {{Evidence
  Based Update}} 2016.
\newblock Brain Stimulation. 2016 Sep-Oct;9(5):641--661.
\newblock doi:{10.1016/j.brs.2016.06.004}.

\bibitem{wilson_clustered_2015}
Wilson D, Moehlis J.
\newblock Clustered {{Desynchronization}} from {{High-Frequency Deep Brain
  Stimulation}}.
\newblock PLOS Computational Biology. 2015;11(12):e1004673.
\newblock doi:{10.1371/journal.pcbi.1004673}.

\bibitem{jiang_irrelevance_2019}
Jiang J, Lai YC.
\newblock Irrelevance of Linear Controllability to Nonlinear Dynamical
  Networks.
\newblock Nature Communications. 2019;10(1):3961.
\newblock doi:{10.1038/s41467-019-11822-5}.

\bibitem{suweis_brain_2019}
Suweis S, Tu C, Rocha RP, Zampieri S, Zorzi M, Corbetta M.
\newblock Brain Controllability: {{Not}} a Slam Dunk Yet.
\newblock NeuroImage. 2019;200:552--555.
\newblock doi:{10.1016/j.neuroimage.2019.07.012}.

\bibitem{tu_warnings_2018}
Tu C, Rocha RP, Corbetta M, Zampieri S, Zorzi M, Suweis S.
\newblock Warnings and Caveats in Brain Controllability.
\newblock NeuroImage. 2018;176:83--91.
\newblock doi:{10.1016/j.neuroimage.2018.04.010}.

\bibitem{gao_target_2014}
Gao J, Liu YY, D'Souza RM, Barab{\'a}si AL.
\newblock Target Control of Complex Networks.
\newblock Nature Communications. 2014;5:5415.
\newblock doi:{10.1038/ncomms6415}.

\bibitem{chen_optimizing_2020}
Chen H, Yong EH.
\newblock Optimizing Target Nodes Selection for the Control Energy of Directed
  Complex Networks.
\newblock Scientific Reports. 2020;10(1):18112.
\newblock doi:{10.1038/s41598-020-75101-w}.

\bibitem{zhao_age-related_2015}
Zhao T, Cao M, Niu H, Zuo XN, Evans A, He Y, et~al.
\newblock Age-Related Changes in the Topological Organization of the White
  Matter Structural Connectome across the Human Lifespan.
\newblock Human Brain Mapping. 2015;36(10):3777--3792.
\newblock doi:{10.1002/hbm.22877}.

\bibitem{gong_age-_2009}
Gong G, {Rosa-Neto} P, Carbonell F, Chen ZJ, He Y, Evans AC.
\newblock Age- and {{Gender-Related Differences}} in the {{Cortical Anatomical
  Network}}.
\newblock Journal of Neuroscience. 2009;29(50):15684--15693.
\newblock doi:{10.1523/JNEUROSCI.2308-09.2009}.

\bibitem{brown_connected_2016}
Brown JA, Van~Horn JD.
\newblock Connected Brains and Minds\textemdash{{The UMCD}} Repository for
  Brain Connectivity Matrices.
\newblock NeuroImage. 2016;124:1238--1241.
\newblock doi:{10.1016/j.neuroimage.2015.08.043}.

\bibitem{bassignana_step-wise_2020}
Bassignana G, Fransson J, Henry V, Colliot O, Zujovic V, De~Vico~Fallani F.
\newblock Stepwise Target Controllability Identifies Dysregulations of
  Macrophage Networks in Multiple Sclerosis.
\newblock Network Neuroscience. 2021;5(2):337--357.
\newblock doi:{10.1162/netn_a_00180}.

\bibitem{zhang_distinct_2019}
Zhang M, Savill N, Margulies DS, Smallwood J, Jefferies E.
\newblock Distinct Individual Differences in Default Mode Network Connectivity
  Relate to Off-Task Thought and Text Memory during Reading.
\newblock Scientific Reports. 2019;9(1):16220.
\newblock doi:{10.1038/s41598-019-52674-9}.

\bibitem{xu_activation_2016}
Xu X, Yuan H, Lei X.
\newblock Activation and {{Connectivity}} within the {{Default Mode Network
  Contribute Independently}} to {{Future-Oriented Thought}}.
\newblock Scientific Reports. 2016;6(1):1--10.
\newblock doi:{10.1038/srep21001}.

\bibitem{davey_exploring_2016}
Davey J, Thompson HE, Hallam G, Karapanagiotidis T, Murphy C, De~Caso I, et~al.
\newblock Exploring the Role of the Posterior Middle Temporal Gyrus in Semantic
  Cognition: {{Integration}} of Anterior Temporal Lobe with Executive
  Processes.
\newblock NeuroImage. 2016;137:165--177.
\newblock doi:{10.1016/j.neuroimage.2016.05.051}.

\bibitem{scheltens_atrophy_1992}
Scheltens P, Leys D, Barkhof F, Huglo D, Weinstein HC, Vermersch P, et~al.
\newblock Atrophy of Medial Temporal Lobes on {{MRI}} in "Probable"
  {{Alzheimer}}'s Disease and Normal Ageing: Diagnostic Value and
  Neuropsychological Correlates.
\newblock Journal of Neurology, Neurosurgery \& Psychiatry.
  1992;55(10):967--972.
\newblock doi:{10.1136/jnnp.55.10.967}.

\bibitem{charras_functional_2015}
Charras P, Herbet G, Deverdun J, {de Champfleur} NM, Duffau H, Bartolomeo P,
  et~al.
\newblock Functional Reorganization of the Attentional Networks in Low-Grade
  Glioma Patients: {{A}} Longitudinal Study.
\newblock Cortex. 2015;63:27--41.
\newblock doi:{10.1016/j.cortex.2014.08.010}.

\bibitem{salvalaggio_post-stroke_2020}
Salvalaggio A, De~Filippo De~Grazia M, Zorzi M, {Thiebaut de Schotten} M,
  Corbetta M.
\newblock Post-Stroke Deficit Prediction from Lesion and Indirect Structural
  and Functional Disconnection.
\newblock Brain. 2020;143(7):2173--2188.
\newblock doi:{10.1093/brain/awaa156}.

\bibitem{corbetta_reorienting_2008}
Corbetta M, Patel G, Shulman GL.
\newblock The {{Reorienting System}} of the {{Human Brain}}: {{From
  Environment}} to {{Theory}} of {{Mind}}.
\newblock Neuron. 2008;58(3):306--324.
\newblock doi:{10.1016/j.neuron.2008.04.017}.

\bibitem{husain_space_2007}
Husain M, Nachev P.
\newblock Space and the Parietal Cortex.
\newblock Trends in Cognitive Sciences. 2007;11(1):30--36.
\newblock doi:{10.1016/j.tics.2006.10.011}.

\bibitem{bartolomeo_hemispheric_2019}
Bartolomeo P, Seidel~Malkinson T.
\newblock Hemispheric Lateralization of Attention Processes in the Human Brain.
\newblock Current Opinion in Psychology. 2019;29:90--96.
\newblock doi:{10.1016/j.copsyc.2018.12.023}.

\bibitem{gotts_two_2013}
Gotts SJ, Jo HJ, Wallace GL, Saad ZS, Cox RW, Martin A.
\newblock Two Distinct Forms of Functional Lateralization in the Human Brain.
\newblock Proceedings of the National Academy of Sciences.
  2013;doi:{10.1073/pnas.1302581110}.

\bibitem{koch_asymmetry_2011}
Koch G, Cercignani M, Bonn{\`i} S, Giacobbe V, Bucchi G, Versace V, et~al.
\newblock Asymmetry of {{Parietal Interhemispheric Connections}} in {{Humans}}.
\newblock Journal of Neuroscience. 2011;31(24):8967--8975.
\newblock doi:{10.1523/JNEUROSCI.6567-10.2011}.

\bibitem{cui_optimization_2020}
Cui Z, Stiso J, Baum GL, Kim JZ, Roalf DR, Betzel RF, et~al.
\newblock Optimization of Energy State Transition Trajectory Supports the
  Development of Executive Function during Youth.
\newblock eLife. 2020;9:e53060.
\newblock doi:{10.7554/eLife.53060}.

\bibitem{medaglia_network_2018}
Medaglia JD, Harvey DY, White N, Kelkar A, Zimmerman J, Bassett DS, et~al.
\newblock Network {{Controllability}} in the {{Inferior Frontal Gyrus Relates}}
  to {{Controlled Language Variability}} and {{Susceptibility}} to {{TMS}}.
\newblock Journal of Neuroscience. 2018;38(28):6399--6410.

\bibitem{honey_can_2010}
Honey CJ, Thivierge JP, Sporns O.
\newblock Can Structure Predict Function in the Human Brain?
\newblock NeuroImage. 2010;52(3):766--776.
\newblock doi:{10.1016/j.neuroimage.2010.01.071}.

\bibitem{narayanan_redundancy_2005}
Narayanan NS, Kimchi EY, Laubach M.
\newblock Redundancy and {{Synergy}} of {{Neuronal Ensembles}} in {{Motor
  Cortex}}.
\newblock The Journal of Neuroscience. 2005;25(17):4207--4216.
\newblock doi:{10.1523/JNEUROSCI.4697-04.2005}.

\bibitem{paquola_microstructural_2019}
Paquola C, Wael RVD, Wagstyl K, Bethlehem RAI, Hong SJ, Seidlitz J, et~al.
\newblock Microstructural and Functional Gradients Are Increasingly Dissociated
  in Transmodal Cortices.
\newblock PLOS Biology. 2019;17(5):e3000284.
\newblock doi:{10.1371/journal.pbio.3000284}.

\bibitem{shafiei_topographic_2020}
Shafiei G, Markello RD, {Vos de Wael} R, Bernhardt BC, Fulcher BD, Misic B.
\newblock Topographic Gradients of Intrinsic Dynamics across Neocortex.
\newblock eLife. 2020;9:e62116.
\newblock doi:{10.7554/eLife.62116}.

\bibitem{wang_macroscopic_2020}
Wang XJ.
\newblock Macroscopic Gradients of Synaptic Excitation and Inhibition in the
  Neocortex.
\newblock Nature Reviews Neuroscience. 2020;21(3):169--178.
\newblock doi:{10.1038/s41583-020-0262-x}.

\bibitem{vazquez-rodriguez_gradients_2019}
{V{\'a}zquez-Rodr{\'i}guez} B, Su{\'a}rez LE, Markello RD, Shafiei G, Paquola
  C, Hagmann P, et~al.
\newblock Gradients of Structure\textendash Function Tethering across
  Neocortex.
\newblock Proceedings of the National Academy of Sciences.
  2019;116(42):21219--21227.
\newblock doi:{10.1073/pnas.1903403116}.

\bibitem{demirtas_hierarchical_2019}
Demirta{\c s} M, Burt JB, Helmer M, Ji JL, Adkinson BD, Glasser MF, et~al.
\newblock Hierarchical {{Heterogeneity}} across {{Human Cortex Shapes
  Large-Scale Neural Dynamics}}.
\newblock Neuron. 2019;101(6):1181--1194.e13.
\newblock doi:{10.1016/j.neuron.2019.01.017}.

\bibitem{sohn_network_2021}
Sohn H, Meirhaeghe N, Rajalingham R, Jazayeri M.
\newblock A {{Network Perspective}} on {{Sensorimotor Learning}}.
\newblock Trends in Neurosciences. 2021;44(3):170--181.
\newblock doi:{10.1016/j.tins.2020.11.007}.

\bibitem{bassett_learning-induced_2015}
Bassett DS, Yang M, Wymbs NF, Grafton ST.
\newblock Learning-Induced Autonomy of Sensorimotor Systems.
\newblock Nature Neuroscience. 2015;18(5):744--751.
\newblock doi:{10.1038/nn.3993}.

\bibitem{adolphs_role_2000}
Adolphs R, Damasio H, Tranel D, Cooper G, Damasio AR.
\newblock A {{Role}} for {{Somatosensory Cortices}} in the {{Visual
  Recognition}} of {{Emotion}} as {{Revealed}} by {{Three-Dimensional Lesion
  Mapping}}.
\newblock Journal of Neuroscience. 2000;20(7):2683--2690.
\newblock doi:{10.1523/JNEUROSCI.20-07-02683.2000}.

\bibitem{kong_sensory-motor_2021}
Kong X, Kong R, Orban C, Wang P, Zhang S, Anderson K, et~al.
\newblock Sensory-Motor Cortices Shape Functional Connectivity Dynamics in the
  Human Brain.
\newblock Nature Communications. 2021;12(1):6373.
\newblock doi:{10.1038/s41467-021-26704-y}.

\bibitem{rolls_limbic_2015}
Rolls ET.
\newblock Limbic Systems for Emotion and for Memory, but No Single Limbic
  System.
\newblock Cortex. 2015;62:119--157.
\newblock doi:{10.1016/j.cortex.2013.12.005}.

\bibitem{catani_revised_2013}
Catani M, Dell'acqua F, {Thiebaut de Schotten} M.
\newblock A Revised Limbic System Model for Memory, Emotion and Behaviour.
\newblock Neuroscience and Biobehavioral Reviews. 2013;37(8):1724--1737.
\newblock doi:{10.1016/j.neubiorev.2013.07.001}.

\bibitem{kaas_evolution_1995}
Kaas JH.
\newblock The {{Evolution}} of {{Isocortex}}.
\newblock Brain, Behavior and Evolution. 1995;46(4-5):187--196.
\newblock doi:{10.1159/000113273}.

\bibitem{rasia-filho_subcortical-allocortical-_2021}
{Rasia-Filho} AA, Guerra KTK, V{\'a}squez CE, Dall'Oglio A, Reberger R, Jung
  CR, et~al.
\newblock The {{Subcortical-Allocortical- Neocortical}} Continuum for the
  {{Emergence}} and {{Morphological Heterogeneity}} of {{Pyramidal Neurons}} in
  the {{Human Brain}}.
\newblock Frontiers in Synaptic Neuroscience. 2021;13:7.
\newblock doi:{10.3389/fnsyn.2021.616607}.

\bibitem{bakkour_effects_2013}
Bakkour A, Morris JC, Wolk DA, Dickerson BC.
\newblock The Effects of Aging and {{Alzheimer}}'s Disease on Cerebral Cortical
  Anatomy: {{Specificity}} and Differential Relationships with Cognition.
\newblock NeuroImage. 2013;76:332--344.
\newblock doi:{10.1016/j.neuroimage.2013.02.059}.

\bibitem{damoiseaux_white_2009}
Damoiseaux JS, Smith SM, Witter MP, {Sanz-Arigita} EJ, Barkhof F, Scheltens P,
  et~al.
\newblock White Matter Tract Integrity in Aging and {{Alzheimer}}'s Disease.
\newblock Human Brain Mapping. 2009;30(4):1051--1059.
\newblock doi:{10.1002/hbm.20563}.

\bibitem{allen_normal_2005}
Allen JS, Bruss J, Brown CK, Damasio H.
\newblock Normal Neuroanatomical Variation Due to Age: {{The}} Major Lobes and
  a Parcellation of the Temporal Region.
\newblock Neurobiology of Aging. 2005;26(9):1245--1260.
\newblock doi:{10.1016/j.neurobiolaging.2005.05.023}.

\bibitem{bernard_evidence_2012}
Bernard JA, Seidler RD.
\newblock Evidence for Motor Cortex Dedifferentiation in Older Adults.
\newblock Neurobiology of Aging. 2012;33(9):1890--1899.
\newblock doi:{10.1016/j.neurobiolaging.2011.06.021}.

\bibitem{voss_dedifferentiation_2008}
Voss MW, Erickson KI, Chaddock L, Prakash RS, Colcombe SJ, Morris KS, et~al.
\newblock Dedifferentiation in the visual cortex: an {fMRI} investigation of
  individual differences in older adults.
\newblock Brain Res. 2008;1244:121--131.
\newblock doi:{10.1016/j.brainres.2008.09.051}.

\bibitem{cabeza_task-independent_2004}
Cabeza R, Daselaar SM, Dolcos F, Prince SE, Budde M, Nyberg L.
\newblock Task-Independent and {{Task-specific Age Effects}} on {{Brain
  Activity}} during {{Working Memory}}, {{Visual Attention}} and {{Episodic
  Retrieval}}.
\newblock Cerebral Cortex. 2004;14(4):364--375.
\newblock doi:{10.1093/cercor/bhg133}.

\bibitem{park_adaptive_2009}
Park DC, {Reuter-Lorenz} P.
\newblock The {{Adaptive Brain}}: {{Aging}} and {{Neurocognitive Scaffolding}}.
\newblock Annual Review of Psychology. 2009;60(1):173--196.
\newblock doi:{10.1146/annurev.psych.59.103006.093656}.

\bibitem{hampson_intrinsic_2012}
Hampson M, Tokoglu F, Shen X, Scheinost D, Papademetris X, Constable RT.
\newblock Intrinsic {{Brain Connectivity Related}} to {{Age}} in {{Young}} and
  {{Middle Aged Adults}}.
\newblock PLOS ONE. 2012;7(9):e44067.
\newblock doi:{10.1371/journal.pone.0044067}.

\bibitem{bagarinao_aging_2020}
Bagarinao E, Watanabe H, Maesawa S, Mori D, Hara K, Kawabata K, et~al.
\newblock Aging {{Impacts}} the {{Overall Connectivity Strength}} of {{Regions
  Critical}} for {{Information Transfer Among Brain Networks}}.
\newblock Frontiers in Aging Neuroscience. 2020;12.
\newblock doi:{10.3389/fnagi.2020.592469}.

\bibitem{bagarinao_reorganization_2019}
Bagarinao E, Watanabe H, Maesawa S, Mori D, Hara K, Kawabata K, et~al.
\newblock Reorganization of Brain Networks and Its Association with General
  Cognitive Performance over the Adult Lifespan.
\newblock Scientific Reports. 2019;9(1):11352.
\newblock doi:{10.1038/s41598-019-47922-x}.

\bibitem{geerligs_brain-wide_2015}
Geerligs L, Renken RJ, Saliasi E, Maurits NM, Lorist MM.
\newblock A {{Brain-Wide Study}} of {{Age-Related Changes}} in {{Functional
  Connectivity}}.
\newblock Cerebral Cortex. 2015;25(7):1987--1999.
\newblock doi:{10.1093/cercor/bhu012}.

\bibitem{betzel_changes_2014}
Betzel RF, Byrge L, He Y, Go{\~n}i J, Zuo XN, Sporns O.
\newblock Changes in Structural and Functional Connectivity among Resting-State
  Networks across the Human Lifespan.
\newblock NeuroImage. 2014;102:345--357.
\newblock doi:{10.1016/j.neuroimage.2014.07.067}.

\bibitem{song_age-related_2014}
Song J, Birn RM, Boly M, Meier TB, Nair VA, Meyerand ME, et~al.
\newblock Age-{{Related Reorganizational Changes}} in {{Modularity}} and
  {{Functional Connectivity}} of {{Human Brain Networks}}.
\newblock Brain Connectivity. 2014;4(9):662--676.
\newblock doi:{10.1089/brain.2014.0286}.

\bibitem{spreng_attenuated_2016}
Spreng RN, Stevens WD, Viviano JD, Schacter DL.
\newblock Attenuated Anticorrelation between the Default and Dorsal Attention
  Networks with Aging: Evidence from Task and Rest.
\newblock Neurobiology of Aging. 2016;45:149--160.
\newblock doi:{10.1016/j.neurobiolaging.2016.05.020}.

\bibitem{tang_developmental_2017}
Tang E, Giusti C, Baum GL, Gu S, Pollock E, Kahn AE, et~al.
\newblock Developmental Increases in White Matter Network Controllability
  Support a Growing Diversity of Brain Dynamics.
\newblock Nature Communications. 2017;8(1):1252.
\newblock doi:{10.1038/s41467-017-01254-4}.

\bibitem{kubicki_early_2016}
Kubicki A, Fautrelle L, Bourrelier J, Rouaud O, Mourey F.
\newblock The {{Early Indicators}} of {{Functional Decrease}} in {{Mild
  Cognitive Impairment}}.
\newblock Frontiers in Aging Neuroscience. 2016;8.

\bibitem{behfar_graph_2020}
Behfar Q, Behfar SK, {von Reutern} B, Richter N, Dronse J, Fassbender R, et~al.
\newblock Graph {{Theory Analysis Reveals Resting-State Compensatory
  Mechanisms}} in {{Healthy Aging}} and {{Prodromal Alzheimer}}'s {{Disease}}.
\newblock Frontiers in Aging Neuroscience. 2020;12.

\bibitem{guillon_disrupted_2019}
Guillon J, Chavez M, Battiston F, Attal Y, La~Corte V, {Thiebaut de Schotten}
  M, et~al.
\newblock Disrupted Core-Periphery Structure of Multimodal Brain Networks in
  {{Alzheimer}}'s Disease.
\newblock Network Neuroscience. 2019;3(2):635--652.

\bibitem{hoffman_age-related_2018}
Hoffman P, Morcom AM.
\newblock Age-Related Changes in the Neural Networks Supporting Semantic
  Cognition: {{A}} Meta-Analysis of 47 Functional Neuroimaging Studies.
\newblock Neuroscience \& Biobehavioral Reviews. 2018;84:134--150.
\newblock doi:{10.1016/j.neubiorev.2017.11.010}.

\bibitem{grady_age_2016}
Grady C, Sarraf S, Saverino C, Campbell K.
\newblock Age Differences in the Functional Interactions among the Default,
  Frontoparietal Control, and Dorsal Attention Networks.
\newblock Neurobiology of Aging. 2016;41:159--172.
\newblock doi:{10.1016/j.neurobiolaging.2016.02.020}.

\bibitem{li_neuromodulation_2014}
Li SC, Rieckmann A.
\newblock Neuromodulation and Aging: Implications of Aging Neuronal Gain
  Control on Cognition.
\newblock Current Opinion in Neurobiology. 2014;29:148--158.
\newblock doi:{10.1016/j.conb.2014.07.009}.

\bibitem{chan_decreased_2014}
Chan MY, Park DC, Savalia NK, Petersen SE, Wig GS.
\newblock Decreased Segregation of Brain Systems across the Healthy Adult
  Lifespan.
\newblock Proceedings of the National Academy of Sciences.
  2014;111(46):E4997--E5006.
\newblock doi:{10.1073/pnas.1415122111}.

\bibitem{de_schotten_direct_2005}
{de Schotten} MT, Urbanski M, Duffau H, Volle E, L{\'e}vy R, Dubois B, et~al.
\newblock Direct {{Evidence}} for a {{Parietal-Frontal Pathway Subserving
  Spatial Awareness}} in {{Humans}}.
\newblock Science. 2005;309(5744):2226--2228.
\newblock doi:{10.1126/science.1116251}.

\bibitem{grady_cognitive_2012}
Grady C.
\newblock The Cognitive Neuroscience of Ageing.
\newblock Nature Reviews Neuroscience. 2012;13(7):491--505.
\newblock doi:{10.1038/nrn3256}.

\bibitem{duda_neurocompensatory_2019}
Duda BM, Owens MM, Hallowell ES, Sweet LH.
\newblock Neurocompensatory {{Effects}} of the {{Default Network}} in {{Older
  Adults}}.
\newblock Frontiers in Aging Neuroscience. 2019;11:111.
\newblock doi:{10.3389/fnagi.2019.00111}.

\bibitem{morcom_neural_2015}
Morcom AM, Johnson W.
\newblock Neural {{Reorganization}} and {{Compensation}} in {{Aging}}.
\newblock Journal of Cognitive Neuroscience. 2015;27(7):1275--1285.

\bibitem{buckner_brains_2008}
Buckner RL, {Andrews-Hanna} JR, Schacter DL.
\newblock The {{Brain}}'s {{Default Network}}.
\newblock Annals of the New York Academy of Sciences. 2008;1124(1):1--38.
\newblock doi:{10.1196/annals.1440.011}.

\bibitem{logan_under-recruitment_2002}
Logan JM, Sanders AL, Snyder AZ, Morris JC, Buckner RL.
\newblock Under-{{Recruitment}} and {{Nonselective Recruitment}}: {{Dissociable
  Neural Mechanisms Associated}} with {{Aging}}.
\newblock Neuron. 2002;33(5):827--840.
\newblock doi:{10.1016/S0896-6273(02)00612-8}.

\bibitem{corbetta_control_2002}
Corbetta M, Shulman GL.
\newblock Control of Goal-Directed and Stimulus-Driven Attention in the Brain.
\newblock Nature Reviews Neuroscience. 2002;3(3):201--215.
\newblock doi:{10.1038/nrn755}.

\bibitem{shulman_right_2010}
Shulman GL, Pope DLW, Astafiev SV, McAvoy MP, Snyder AZ, Corbetta M.
\newblock Right {{Hemisphere Dominance}} during {{Spatial Selective Attention}}
  and {{Target Detection Occurs Outside}} the {{Dorsal Frontoparietal
  Network}}.
\newblock Journal of Neuroscience. 2010;30(10):3640--3651.
\newblock doi:{10.1523/JNEUROSCI.4085-09.2010}.

\bibitem{murray_attention_2004}
Murray SO, Wojciulik E.
\newblock Attention Increases Neural Selectivity in the Human Lateral Occipital
  Complex.
\newblock Nature Neuroscience. 2004;7(1):70--74.
\newblock doi:{10.1038/nn1161}.

\bibitem{cabeza_hemispheric_2002}
Cabeza R.
\newblock Hemispheric Asymmetry Reduction in Older Adults: The {{HAROLD}}
  Model.
\newblock Psychology and Aging. 2002;17(1):85--100.
\newblock doi:{10.1037//0882-7974.17.1.85}.

\bibitem{tomasi_aging_2012}
Tomasi D, Volkow ND.
\newblock Aging and Functional Brain Networks.
\newblock Molecular Psychiatry. 2012;17(5):549--558.
\newblock doi:{10.1038/mp.2011.81}.

\bibitem{spreng_shifting_2019}
Spreng RN, Turner GR.
\newblock The {{Shifting Architecture}} of {{Cognition}} and {{Brain Function}}
  in {{Older Adulthood}}.
\newblock Perspectives on Psychological Science. 2019;14(4):523--542.
\newblock doi:{10.1177/1745691619827511}.

\bibitem{medaglia_brain_2017}
Medaglia JD, Pasqualetti F, Hamilton RH, {Thompson-Schill} SL, Bassett DS.
\newblock Brain and Cognitive Reserve: {{Translation}} via Network Control
  Theory.
\newblock Neuroscience \& Biobehavioral Reviews. 2017;75:53--64.
\newblock doi:{10.1016/j.neubiorev.2017.01.016}.

\bibitem{cheng_reorganization_2012}
Cheng L, Wu Z, Fu Y, Miao F, Sun J, Tong S.
\newblock Reorganization of Functional Brain Networks during the Recovery of
  Stroke: {{A}} Functional {{MRI}} Study.
\newblock In: 2012 {{Annual International Conference}} of the {{IEEE
  Engineering}} in {{Medicine}} and {{Biology Society}}; 2012. p. 4132--4135.

\bibitem{zhu_disrupted_2017}
Zhu Y, Bai L, Liang P, Kang S, Gao H, Yang H.
\newblock Disrupted Brain Connectivity Networks in Acute Ischemic Stroke
  Patients.
\newblock Brain Imaging and Behavior. 2017;11(2):444--453.
\newblock doi:{10.1007/s11682-016-9525-6}.

\bibitem{mancuso_homotopic_2019}
Mancuso L, Costa T, Nani A, Manuello J, Liloia D, Gelmini G, et~al.
\newblock The Homotopic Connectivity of the Functional Brain: A Meta-Analytic
  Approach.
\newblock Scientific Reports. 2019;9(1):3346.
\newblock doi:{10.1038/s41598-019-40188-3}.

\bibitem{tang_decreased_2016}
Tang C, Zhao Z, Chen C, Zheng X, Sun F, Zhang X, et~al.
\newblock Decreased {{Functional Connectivity}} of {{Homotopic Brain Regions}}
  in {{Chronic Stroke Patients}}: {{A Resting State fMRI Study}}.
\newblock PLOS ONE. 2016;11(4):e0152875.
\newblock doi:{10.1371/journal.pone.0152875}.

\bibitem{grefkes_connectivity-based_2014}
Grefkes C, Fink GR.
\newblock Connectivity-Based Approaches in Stroke and Recovery of Function.
\newblock The Lancet Neurology. 2014;13(2):206--216.
\newblock doi:{10.1016/S1474-4422(13)70264-3}.

\bibitem{buetefisch_role_2015}
Buetefisch CM.
\newblock Role of the {{Contralesional Hemisphere}} in {{Post-Stroke Recovery}}
  of {{Upper Extremity Motor Function}}.
\newblock Frontiers in Neurology. 2015;6:214.
\newblock doi:{10.3389/fneur.2015.00214}.

\bibitem{bartolomeo_let_2016}
Bartolomeo P, {Thiebaut de Schotten} M.
\newblock Let Thy Left Brain Know What Thy Right Brain Doeth:
  {{Inter-hemispheric}} Compensation of Functional Deficits after Brain Damage.
\newblock Neuropsychologia. 2016;93:407--412.
\newblock doi:{10.1016/j.neuropsychologia.2016.06.016}.

\bibitem{bartolomeo_competition_2021}
Bartolomeo P.
\newblock From Competition to Cooperation: {{Visual}} Neglect across the
  Hemispheres.
\newblock Revue Neurologique. 2021;177(9):1104--1111.
\newblock doi:{10.1016/j.neurol.2021.07.015}.

\bibitem{kullmann_editorial_2019}
Kullmann DM.
\newblock Editorial.
\newblock Brain. 2019;142(4):833.
\newblock doi:{10.1093/brain/awz077}.

\bibitem{reich_independent_2001}
Reich DS, Mechler F, Victor JD.
\newblock Independent and {{Redundant Information}} in {{Nearby Cortical
  Neurons}}.
\newblock Science. 2001;294(5551):2566--2568.
\newblock doi:{10.1126/science.1065839}.

\bibitem{so_redundant_2012}
So K, Ganguly K, Jimenez J, Gastpar MC, Carmena JM.
\newblock Redundant Information Encoding in Primary Motor Cortex during Natural
  and Prosthetic Motor Control.
\newblock Journal of Computational Neuroscience. 2012;32(3):555--561.
\newblock doi:{10.1007/s10827-011-0369-1}.

\bibitem{griffis_structural_2019}
Griffis JC, Metcalf NV, Corbetta M, Shulman GL.
\newblock Structural {{Disconnections Explain Brain Network Dysfunction}} after
  {{Stroke}}.
\newblock Cell Reports. 2019;28(10):2527--2540.e9.
\newblock doi:{10.1016/j.celrep.2019.07.100}.

\bibitem{betzel_optimally_2016}
Betzel RF, Gu S, Medaglia JD, Pasqualetti F, Bassett DS.
\newblock Optimally Controlling the Human Connectome: The Role of Network
  Topology.
\newblock Scientific Reports. 2016;6(1):30770.
\newblock doi:{10.1038/srep30770}.

\bibitem{dosenbach_prediction_2010}
Dosenbach NUF, Nardos B, Cohen AL, Fair DA, Power JD, Church JA, et~al.
\newblock Prediction of {{Individual Brain Maturity Using fMRI}}.
\newblock Science. 2010;329(5997):1358--1361.
\newblock doi:{10.1126/science.1194144}.

\bibitem{fair_functional_2009}
Fair DA, Cohen AL, Power JD, Dosenbach NUF, Church JA, Miezin FM, et~al.
\newblock Functional {{Brain Networks Develop}} from a ``{{Local}} to
  {{Distributed}}'' {{Organization}}.
\newblock PLOS Computational Biology. 2009;5(5):e1000381.
\newblock doi:{10.1371/journal.pcbi.1000381}.

\bibitem{supekar_developmental_2012}
Supekar K, Menon V.
\newblock Developmental {{Maturation}} of {{Dynamic Causal Control Signals}} in
  {{Higher-Order Cognition}}: {{A Neurocognitive Network Model}}.
\newblock PLOS Computational Biology. 2012;8(2):e1002374.
\newblock doi:{10.1371/journal.pcbi.1002374}.

\bibitem{giza_is_2006}
Giza CC, Prins ML.
\newblock Is {{Being Plastic Fantastic}}? {{Mechanisms}} of {{Altered
  Plasticity}} after {{Developmental Traumatic Brain Injury}}.
\newblock Developmental Neuroscience. 2006;28(4-5):364--379.
\newblock doi:{10.1159/000094163}.

\bibitem{kornfeld_cortical_2015}
Kornfeld S, Delgado~Rodr{\'i}guez JA, Everts R, {Kaelin-Lang} A, Wiest R,
  Weisstanner C, et~al.
\newblock Cortical Reorganisation of Cerebral Networks after Childhood Stroke:
  Impact on Outcome.
\newblock BMC Neurology. 2015;15(1):90.
\newblock doi:{10.1186/s12883-015-0309-1}.

\bibitem{anderson_children_2011}
Anderson V, {Spencer-Smith} M, Wood A.
\newblock Do Children Really Recover Better? {{Neurobehavioural}} Plasticity
  after Early Brain Insult.
\newblock Brain. 2011;134(8):2197--2221.
\newblock doi:{10.1093/brain/awr103}.

\bibitem{max_pediatric_2010}
Max JE, Bruce M, Keatley E, Delis D.
\newblock Pediatric {{Stroke}}: {{Plasticity}}, {{Vulnerability}}, and {{Age}}
  of {{Lesion Onset}}.
\newblock The Journal of Neuropsychiatry and Clinical Neurosciences.
  2010;22(1):30--39.
\newblock doi:{10.1176/jnp.2010.22.1.30}.

\bibitem{dennis_age_2013}
Dennis M, Spiegler BJ, Juranek JJ, Bigler ED, Snead OC, Fletcher JM.
\newblock Age, Plasticity, and Homeostasis in Childhood Brain Disorders.
\newblock Neuroscience \& Biobehavioral Reviews. 2013;37(10, Part
  2):2760--2773.
\newblock doi:{10.1016/j.neubiorev.2013.09.010}.

\bibitem{liu_control_2016}
Liu YY, Barab{\'a}si AL.
\newblock Control {{Principles}} of {{Complex Networks}}.
\newblock Reviews of Modern Physics. 2016;88(3).
\newblock doi:{10.1103/RevModPhys.88.035006}.

\bibitem{slotine_applied_1991}
Slotine JJ, Li W.
\newblock Applied {{Nonlinear Control}}.
\newblock {Englewood Cliffs, N.J}: {Pearson}; 1991.

\bibitem{kerepesi_how_2016}
Kerepesi C, Szalkai B, Varga B, Grolmusz V.
\newblock How to {{Direct}} the {{Edges}} of the {{Connectomes}}: {{Dynamics}}
  of the {{Consensus Connectomes}} and the {{Development}} of the
  {{Connections}} in the {{Human Brain}}.
\newblock PLOS ONE. 2016;11(6):e0158680.
\newblock doi:{10.1371/journal.pone.0158680}.

\bibitem{seguin_inferring_2019}
Seguin C, Razi A, Zalesky A.
\newblock Inferring Neural Signalling Directionality from Undirected Structural
  Connectomes.
\newblock Nature Communications. 2019;10(1):1--13.
\newblock doi:{10.1038/s41467-019-12201-w}.

\bibitem{worrell_optimized_2017}
Worrell JC, Rumschlag J, Betzel RF, Sporns O, Mi{\v s}i{\'c} B.
\newblock Optimized Connectome Architecture for Sensory-Motor Integration.
\newblock Network Neuroscience. 2017;1(4):415--430.

\bibitem{goni_exploring_2013}
Go{\~n}i J, {Avena-Koenigsberger} A, {Velez de Mendizabal} N, {van den Heuvel}
  MP, Betzel RF, Sporns O.
\newblock Exploring the {{Morphospace}} of {{Communication Efficiency}} in
  {{Complex Networks}}.
\newblock PLoS ONE. 2013;8(3):e58070.
\newblock doi:{10.1371/journal.pone.0058070}.

\bibitem{avena-koenigsberger_communication_2018}
{Avena-Koenigsberger} A, Misic B, Sporns O.
\newblock Communication Dynamics in Complex Brain Networks.
\newblock Nature Reviews Neuroscience. 2018;19(1):17--33.
\newblock doi:{10.1038/nrn.2017.149}.

\bibitem{raj_network_2012}
Raj A, Kuceyeski A, Weiner M.
\newblock A {{Network Diffusion Model}} of {{Disease Progression}} in
  {{Dementia}}.
\newblock Neuron. 2012;73(6):1204--1215.
\newblock doi:{10.1016/j.neuron.2011.12.040}.

\bibitem{abdelnour_functional_2018}
Abdelnour F, Dayan M, Devinsky O, Thesen T, Raj A.
\newblock Functional Brain Connectivity Is Predictable from Anatomic Network's
  {{Laplacian}} Eigen-Structure.
\newblock NeuroImage. 2018;172:728--739.
\newblock doi:{10.1016/j.neuroimage.2018.02.016}.

\bibitem{dubois_single-unit_2015}
Dubois J, de~Berker AO, Tsao DY.
\newblock Single-{{Unit Recordings}} in the {{Macaque Face Patch System Reveal
  Limitations}} of {{fMRI MVPA}}.
\newblock Journal of Neuroscience. 2015;35(6):2791--2802.
\newblock doi:{10.1523/JNEUROSCI.4037-14.2015}.

\bibitem{schilling_histological_2018}
Schilling KG, Janve V, Gao Y, Stepniewska I, Landman BA, Anderson AW.
\newblock Histological Validation of Diffusion {{MRI}} Fiber Orientation
  Distributions and Dispersion.
\newblock NeuroImage. 2018;165:200--221.
\newblock doi:{10.1016/j.neuroimage.2017.10.046}.

\bibitem{huntenburg_large-scale_2018}
Huntenburg JM, Bazin PL, Margulies DS.
\newblock Large-{{Scale Gradients}} in {{Human Cortical Organization}}.
\newblock Trends in Cognitive Sciences. 2018;22(1):21--31.
\newblock doi:{10.1016/j.tics.2017.11.002}.

\bibitem{malkinson_perception_2021}
Malkinson TS, Bayle DJ, Bourgeois A, Lehongre K, {Fernandez-Vidal} S, Navarro
  V, et~al.
\newblock From Perception to Action: {{Intracortical}} Recordings Reveal
  Cortical Gradients of Human Exogenous Attention.
\newblock biorxiv. 2021;doi:{10.1101/2021.01.02.425103}.

\bibitem{nooner_nki-rockland_2012}
Nooner KB, Colcombe S, Tobe R, Mennes M, Benedict M, Moreno A, et~al.
\newblock The {{NKI-Rockland Sample}}: {{A Model}} for {{Accelerating}} the
  {{Pace}} of {{Discovery Science}} in {{Psychiatry}}.
\newblock Frontiers in Neuroscience. 2012;6.
\newblock doi:{10.3389/fnins.2012.00152}.

\bibitem{brown_ucla_2012}
Brown J, Rudie J, Bandrowski A, Van~Horn J, Bookheimer S.
\newblock The {{UCLA}} Multimodal Connectivity Database: A Web-Based Platform
  for Brain Connectivity Matrix Sharing and Analysis.
\newblock Frontiers in Neuroinformatics. 2012;6.

\bibitem{craddock_whole_2012}
Craddock RC, James GA, Holtzheimer PE, Hu XP, Mayberg HS.
\newblock A Whole Brain {{fMRI}} Atlas Generated via Spatially Constrained
  Spectral Clustering.
\newblock Human Brain Mapping. 2012;33(8):1914--1928.
\newblock doi:{10.1002/hbm.21333}.

\bibitem{gomez-gardenes_entropy_2008}
{G{\'o}mez-Garde{\~n}es} J, Latora V.
\newblock Entropy Rate of Diffusion Processes on Complex Networks.
\newblock Physical Review E. 2008;78(6):065102.
\newblock doi:{10.1103/PhysRevE.78.065102}.

\bibitem{tewarie_minimum_2015}
Tewarie P, {van Dellen} E, Hillebrand A, Stam CJ.
\newblock The Minimum Spanning Tree: {{An}} Unbiased Method for Brain Network
  Analysis.
\newblock NeuroImage. 2015;104:177--188.
\newblock doi:{10.1016/j.neuroimage.2014.10.015}.

\bibitem{van_dellen_minimum_2018}
{van Dellen} E, Sommer IE, Bohlken MM, Tewarie P, Draaisma L, Zalesky A, et~al.
\newblock Minimum Spanning Tree Analysis of the Human Connectome.
\newblock Human Brain Mapping. 2018;39(6):2455--2471.
\newblock doi:{10.1002/hbm.24014}.

\bibitem{liu_controllability_2011}
Liu YY, Slotine JJ, Barab{\'a}si AL.
\newblock Controllability of Complex Networks.
\newblock Nature. 2011;473(7346):167--173.
\newblock doi:{10.1038/nature10011}.

\bibitem{kalman_mathematical_1963}
Kalman RE.
\newblock Mathematical {{Description}} of {{Linear Dynamical Systems}}.
\newblock Journal of the Society for Industrial and Applied Mathematics Series
  A Control. 1963;1(2):152--192.
\newblock doi:{10.1137/0301010}.

\bibitem{rugh_linear_1995}
Rugh WJ, Kailath T.
\newblock Linear {{System Theory}}, 2nd {{Edition}}.
\newblock 2nd ed. {Upper Saddle River, NJ}: {Pearson}; 1995.

\bibitem{van_den_heuvel_network_2013}
{van den Heuvel} MP, Sporns O.
\newblock Network Hubs in the Human Brain.
\newblock Trends in Cognitive Sciences. 2013;17(12):683--696.
\newblock doi:{10.1016/j.tics.2013.09.012}.

\bibitem{de_vico_fallani_graph_2014}
De~Vico~Fallani F, Richiardi J, Chavez M, Achard S.
\newblock Graph analysis of functional brain networks: practical issues in
  translational neuroscience.
\newblock Phil Trans R Soc B. 2014;369(1653):20130521.
\newblock doi:{10.1098/rstb.2013.0521}.

\bibitem{benjamini_controlling_1995}
Benjamini Y, Hochberg Y.
\newblock Controlling the {{False Discovery Rate}}: {{A Practical}} and
  {{Powerful Approach}} to {{Multiple Testing}}.
\newblock Journal of the Royal Statistical Society Series B (Methodological).
  1995;57(1):289--300.

\end{thebibliography}
\end{small}

\newpage

\newpage

\newpage

\newpage

\newpage

\par\null


\end{document}